\documentclass[fleqn,usenatbib]{mnras}

\usepackage{newtxtext,newtxmath}

\usepackage[T1]{fontenc}
\usepackage{ae,aecompl}

\usepackage{graphicx}	
\usepackage{amsmath}	
\usepackage{amssymb}	
\usepackage{pdflscape}
\usepackage{rotating}

\newcommand{\eg}{e.g.,\ }

\newcommand{\Msun}{$M_{\odot}$}

\newcommand{\kms}{km\,s$^{-1}$}

\newcommand{\OI}{O~{\sc i}}

\newcommand{\MgII}{Mg~{\sc ii}}

\newcommand{\SiII}{Si~{\sc ii}}

\newcommand{\CaII}{Ca~{\sc ii}}

\newcommand{\FeII}{Fe~{\sc ii}}

\newcommand{\CoII}{Co~{\sc ii}}

\newcommand{\NiII}{Ni~{\sc ii}}

\newcommand{\Fefs}{$^{56}$Fe}
\newcommand{\Cofs}{$^{56}$Co}
\newcommand{\Nifs}{$^{56}$Ni}
\newcommand{\Mej}{$M_{\rm ej}$}
\newcommand{\KE}{$E_{\rm kin}$}

\newcommand{\ab}{$\sim$}

\title[GRB\,161219B \& SN\,2016jca]{GRB\,161219B/SN\,2016jca: a powerful stellar collapse }

\author[C. Ashall et al.]{
\parbox{\textwidth}{
\raggedright
C. Ashall$^{1,2}$, $\thanks{E-mail:chris.ashall24@gmail.com}$
P.A. Mazzali$^{2,3}$, 
E. Pian$^{4}$, 
S.E. Woosley$^{5}$,
E. Palazzi$^{4}$, 
S.J. Prentice$^{2,6}$, 
S. Kobayashi$^{2}$, 
S. Holmbo$^{7}$,
A. Levan$^{8}$, 
D. Perley$^{2}$, 
M.D. Stritzinger$^{7}$,
F. Bufano$^{9}$,
A.V. Filippenko$^{10,11}$,
A. Melandri$^{12}$,
S. Oates$^{8}$,  
A. Rossi$^{4}$,
J. Selsing$^{13}$,
W. Zheng$^{10}$, 
A.J. Castro-Tirado$^{14}$,
G. Chincarini$^{12}$,
P. D'Avanzo$^{12}$,
M. De Pasquale$^{15}$, 
S. Emery$^{16}$,
A.S. Fruchter$^{17}$,
K. Hurley$^{18}$,
P. Moller$^{19}$,
K. Nomoto$^{20}$, 
M. Tanaka$^{21}$,
A.F. Valeev$^{22}$\\
}\vspace{0.4cm}\\
\parbox{\textwidth}{
$^{1}$ Department of Physics, Florida State University, Tallahassee, FL 32306, USA\\
$^{2}$ Astrophysics Research Institute, Liverpool John Moores University, IC2, Liverpool Science Park, 146 Brownlow Hill, Liverpool L3 5RF, UK\\
$^{3}$ Max-Planck-Institut f\"ur Astrophysik, Karl-Schwarzschild-Str. 1, D-85748 Garching, Germany\\
$^{4}$ INAF-OAS Bologna, Via P. Gobetti 93/3, 40129 Bologna, Italy\\
$^{5}$ Department of Astronomy and Astrophysics, University of California,
Santa Cruz, CA 95064, USA\\
$^{6}$ Astrophysics Research Centre, School of Mathematics and Physics, Queen's University Belfast, BT7 1NN, UK\\
$^{7}$   Department of Physics and Astronomy, Aarhus University, DK-8000 Aarhus C, Denmark\\
$^{8}$  Department of Physics, University of Warwick, Coventry, CV4 7AL, UK\\
$^{9}$  INAF, Astronomical Observatory of Catania, Italy\\
$^{10}$ Department of Astronomy, University of California, Berkeley, CA  94720-3411, USA\\
$^{11}$ Miller Senior Fellow, Miller Institute for Basic Research in Science, University of California, Berkeley, CA  94720, USA\\
$^{12}$ INAF,  Brera Astronomical Observatory, Italy\\
$^{13}$ Dark Cosmology Centre, Niels Bohr Institute, University of Copenhagen, Juliane Maries Vej 30, 2100 K{\o}benhavn {\o}, Denmark \\
$^{14}$ Instituto de Astrofisica de Andalucia (IAA-CSIC), Glorieta de la Astronomia, E-18008 Granada, Spain\\
$^{15}$ Department of Astronomy and Space Sciences, Istanbul University Beyazid, Istanbul 34119, Turkey\\
$^{16}$ Mullard Space Science Laboratory, University College London, Holmbury St. Mary, Dorking, Surrey RH5 6NT, UK \\
$^{17}$  Space Telescope Science Institute, 3700 San Martin Dr.,  Baltimore, MD 20218, USA \\
$^{18}$ University of California, Berkeley, Space Sciences Laboratory, 7 Gauss Way, Berkeley, CA 94720-7450, USA\\
$^{19}$  European Southern Observatory, Karl-Schwarzschild-Str. 2, 85748 Garching bei M\"{u}nchen, Germany \\
$^{20}$ Kavli Institute for the Physics and Mathematics of the Universe (WPI), The University of Tokyo, Kashiwa, Chiba 277-8583, Japan \\
$^{21}$ National Astronomical Observatory of Japan, Mitaka, Tokyo 181-8588, Japan \\
$^{22}$ Special Astrophysical Observatory, Nizhnij Arkhyz, Karachai-Cherkessian Republic, 369167 Russia\\
}
\vspace{-0.75cm}
}
\date{Accepted XXX. Received \today; in original form ZZZ}

\pubyear{2018}

\begin{document}
\label{firstpage}

\pagerange{\pageref{firstpage}--\pageref{lastpage}}
\maketitle

\begin{abstract}
We report observations and analysis of the nearby gamma-ray burst GRB\,161219B (redshift $z=0.1475$)
and the associated Type Ic supernova (SN) 2016jca.  
GRB\,161219B had an isotropic gamma-ray energy of $\sim 1.6 \times 10^{50}$\,erg. 
Its afterglow is likely refreshed at an epoch preceding the first photometric points (0.6\,d), which slows down the decay rates.
Combined analysis of the SN light curve and multiwavelength observations of the afterglow suggest that the GRB jet was broad during the afterglow phase (full 
opening angle $\sim 42^\circ \pm 3^\circ$).  
Our spectral series shows broad absorption lines typical of GRB supernovae (SNe), which testify to 
the presence of material with velocities up to $\sim 0.25$c. The spectrum at
3.73\,d allows for the very early identification of a SN associated with a
GRB.  Reproducing it requires a large photospheric velocity ($35,000 \pm 7000$\,\kms).
The kinetic energy of the SN is estimated through models to be
\KE $\approx 4 \times 10^{52}$\,erg in spherical symmetry.
The ejected mass in the explosion was \Mej $\approx 6.5 \pm 1.5$\,\Msun,
much less than that of other GRB-SNe, demonstrating diversity among these events. 
The total amount of \Nifs\ in the explosion was $0.27 \pm 0.05$\,\Msun.
The observed spectra require the presence of freshly
synthesised \Nifs\ at the highest velocities, at least 3 times more than a standard GRB-SN.
We also find evidence for a decreasing \Nifs\ abundance as a function of decreasing velocity. 
This suggests that SN\,2016jca was a 
highly aspherical explosion viewed close to on-axis, powered by a compact remnant.
Applying a typical correction for asymmetry, the energy of SN\,2016jca was 
 $\sim$ (1--3) $\times 10^{52}$\,erg, confirming that most of the energy produced by
GRB-SNe goes into the kinetic energy of the SN ejecta.  
\end{abstract}

\begin{keywords}
gamma-ray burst --- supernova
\end{keywords}



\section{Introduction}

When massive stars, with helium cores smaller than 65\,\Msun, exhaust their nuclear fuel their cores collapse to a 
compact object (a neutron star or a black hole; \citealt{Woosley17}). The collapse triggers a supernova (SN)
explosion, in which the outer layers of the star are expelled and a luminous
display is created. If the SN is of Type I, the main power source for its luminosity is usually the 
radioactive decay of \Nifs, which is synthesised in the stellar layers above the 
compact remnant, or along the jet axis, and ejected in the SN \citep{Barnes18}.
However, circumstellar interaction, magnetar \citep{Kasen10,Woosley10} energy input, and black hole accretion \citep{Woosley93,MacFadyen99,Dexter13} may be important in some cases.
A particular subgroup of core-collapse supernovae (SNe) are linked to long-duration gamma-ray 
bursts (GRBs; these include X-ray flashes, a soft-spectrum variety of GRBs).  
Although the connection is well established, it is not well understood. 

Long-duration GRB-SNe (henceforth, simply GRB-SNe) are all of Type Ic \citep{Filippenko97,Woosley06} --- they are produced by stars
that have lost their outer hydrogen and helium layers.   
They  are distinct from other
SNe\,Ic in having broad lines (SNe~Ic-BL), indicating high kinetic energy (isotropic \KE\ $\approx 5 \times 10^{52}$\,erg),
which should be corrected downward by a factor 3--5 to account for asphericity
\citep{Maeda02}).  Those related to classical GRBs (as opposed to X-ray flashes or outbursts)  have relatively high luminosities \citep{Mazzali14,Lyman16,Prentice16,Prentice19}, producing $\sim 0.3$--0.5\,\Msun\ of \Nifs, whose radioactive decay into \Cofs\ and then \Fefs\
powers their light curves \citep{Mazzali00,Drout11}. GRB-SNe are also the most massive
SNe\,Ic. They have typical ejected masses of $\sim 10$\,\Msun, suggesting that the
progenitor stars had initial masses of 30--50\,\Msun\
\citep{Iwamoto98,Deng05,Mazzali13}. Stars in this mass range can collapse to
neutron stars or black holes \citep{Ugliano12}, but the energy of GRB-SNe is far
larger than what the classical neutrino-driven mechanism is likely to achieve \citep{Janka12}. The
source of the high SN \KE\ is therefore likely to be the compact remnant.

Fewer than a dozen SNe connected with high-energy events (GRBs, X-ray flashes and outbursts), all at redshift $z < 0.2$, have accurate photometric and spectroscopic time-series \citep{Galama98,Hjorth03,Stanek03,Matheson03,Ferrero06,Mazzali08a,Tanaka09,Bufano2012,Izzo19,Delia15,Toy16}). 
GRB\,161219B \citep{GCN20296}, which had an observed isotropic energy in gamma-rays of $1.6 \times 10^{50}$\,erg \citep{Frederiks17}, 
exploded on 2016 Dec. 19 (UT dates are used throughout this paper)
in a galaxy at $z=0.1475$ \citep{Tanvir16,Cano17}. 
About 5\,d after the GRB exploded a SN component was detected underlying the
optical afterglow.   Its spectral features were typical of those of
previously observed GRB-SNe, warranting a classification as an SN~Ic \citep{Pian17TNS},
specifically ``SN~Ic-3" \citep{Prentice17}.   \citet{Cano17} reported the presence of a pre-maximum bump that is reminiscent of the early-time behaviour of other SNe, both accompanied and not accompanied by a GRB.  Here we report  on optical photometric and 
spectroscopic follow-up observations of the afterglow of GRB\,161219B and the associated SN\,2016jca, and compare  them with X-ray and ultraviolet (UV) observations made with the Neil Gehrels {\it Swift}  Observatory.  
We also present radiative-transfer spectral models of the SN. 

\section{Observations, data reduction and results}

Optical photometric and spectroscopic observations of the counterpart of  GRB\,161219B commenced on December 22.08, or 1.99\,d  after explosion in the rest frame.   Thirteen spectra of the point-like optical transient were obtained with the
VLT between 1.99 and 268\,d (rest frame) after the GRB, and $BVRI$ photometry was
acquired in various optical filters in the same time interval.   The transient was also observed in {\it BVgri} using IO:O on the 2\,m Liverpool Telescope \citep[LT;][]{Steele04} and in $BVRI$ using the DOLORES camera on the Italian 3.6\,m Telescopio Nazionale Galileo (TNG) at Observatorio del Roque de Los Muchachos, Spain. Photometry was also obtained with the Low Resolution Imaging Spectrometer \citep[LRIS;][]{Oke95} with the 
10m Keck-1 telescope on Maunakea, Hawaii, USA. A log of the photometric and spectroscopic observations can be found in Tables \ref{table:logofphot} and \ref{table:logofspectra}, respectively.   In addition, we downloaded the {\it Swift} XRT data of the GRB counterpart from the archive and re-analysed the {\it Swift} UVOT observations presented by \citet{Cano17}.

\begin{table*}
 \centering
 \caption{Log of photometric observations (mag).} 
 \small
  \begin{tabular}{ccccccccc}
  \hline
MJD & $B$ & $V$ & $R$ & $I$ & $g$ & $r$ & $i$ & Telescope\\
  \hline
57744.08 &  $19.83 \pm{0.01}$  &  $19.66 \pm{0.01}$  &  $19.39 \pm{0.01}$  &  $18.89 \pm{0.00}$  &  -  &  -  &  -  &  VLT \\
57746.08 &  $20.04 \pm{0.01}$  &  $19.84 \pm{0.01}$  &  $19.74 \pm{0.01}$  &  $19.33 \pm{0.01}$  &  -  &  -  &  -  &  VLT \\
57748.12 &  $20.24 \pm{0.01}$  &  $19.82 \pm{0.01}$  &  $19.73 \pm{0.01}$  &  $19.48 \pm{0.01}$  &  -  &  -  &  -  &  VLT \\
57749.03 &  $20.21 \pm{0.03}$ & $19.65 \pm{0.10}$ & $19.53\pm{0.02}$ & $19.35\pm{0.05}$&-&-&-&TNG\\
57749.99 &  $20.18 \pm{0.13}$  &  $19.8 \pm{0.2}$  &  -  &  -  &  $19.81 \pm{0.05}$  &  $19.63 \pm{0.07}$  &  $19.80 \pm{0.03}$  &  LT \\
57750.00 &  -  &  $19.8 \pm{0.2}$  &  -  &  -  &  $19.80 \pm{0.05}$  &  $19.63 \pm{0.07}$  &  $19.80 \pm{0.03}$  &  LT \\
57751.09 &  $20.46 \pm{0.01}$  &  $19.79 \pm{0.01}$  &  $19.64 \pm{0.01}$  &  $19.44 \pm{0.01}$  &  -  &  -  &  -  &  VLT \\
57751.98 &  -  &  $19.63 \pm{0.05}$  &  -  &  -  &  $19.92 \pm{0.05}$  &  $19.54 \pm{0.09}$  &  $19.76 \pm{0.08}$  &  LT \\
57751.99 &  -  &  $19.63 \pm{0.05}$  &  -  &  -  &  $19.92 \pm{0.05}$  &  $19.54 \pm{0.09}$  &  $19.75 \pm{0.08}$  &  LT \\
57753.99& $20.52\pm{0.21}$&$19.77\pm{0.06}$&$19.43\pm{0.06}$&$19.35\pm{0.12}$&-&-&-&TNG\\
57754.28 &  $20.57 \pm{0.01}$  &  $19.82 \pm{0.01}$  &  $19.53 \pm{0.01}$  &  $19.40 \pm{0.01}$  &  -  &  -  &  -  &  VLT \\
57756.36 &  $20.87 \pm{0.01}$  &  $20.17 \pm{0.03}$  &  $19.64 \pm{0.03}$  &  $19.44 \pm{0.04}$  &  -  &  -  &  -  &  Keck \\
57757.00 &  $20.60 \pm{0.09}$  &  $20.2 \pm{0.1}$  &  -  &  -  &  $20.28 \pm{0.03}$  &  $19.64 \pm{0.03}$  &  $19.76 \pm{0.02}$  &  LT \\
57757.01 &  $20.60 \pm{0.09}$  &  $20.2 \pm{0.1}$  &  -  &  -  &  $20.28 \pm{0.03}$  &  $19.64 \pm{0.03}$  &  $19.76 \pm{0.02}$  &  LT \\
57757.98 &  $20.90 \pm{0.06}$  &  $20.03 \pm{0.03}$  &  -  &  -  &  $20.39 \pm{0.04}$  &  $19.77 \pm{0.07}$  &  $19.76 \pm{0.05}$  &  LT \\
57757.98 &  $20.90 \pm{0.06}$  &  $20.03 \pm{0.03}$  &  -  &  -  &  $20.38 \pm{0.04}$  &  $19.75 \pm{0.08}$  &  $19.75 \pm{0.05}$  &  LT \\
57758.04 &  $20.92 \pm{0.02}$  &  $19.96 \pm{0.01}$  &  $19.58 \pm{0.01}$  &  $19.41 \pm{0.01}$  &  -  &  -  &  -  &  VLT \\
57758.96 &  -  &  -  &  -  &  -  &  $20.44 \pm{0.08}$  &  $19.669 \pm{0.08}$  &  $19.59 \pm{0.06}$  &  LT \\
57760.06 &  $21.18 \pm{0.02}$  &  $20.12 \pm{0.01}$  &  $19.66 \pm{0.01}$  &  $19.47 \pm{0.01}$  &  -  &  -  &  -  &  VLT \\
57763.26 &  $21.58 \pm{0.06}$  &  $20.33 \pm{0.02}$  &  $19.73 \pm{0.01}$  &  $19.47 \pm{0.01}$  &  -  &  -  &  -  &  VLT \\
57767.17 &  -  &  $20.51 \pm{0.03}$  &  $19.97 \pm{0.02}$  &  $19.61 \pm{0.02}$  &  -  &  -  &  -  &  VLT \\
57769.98 &  -  &  -  &  -  &  -  &  -  &  $20.35 \pm{0.05}$  &  $20.05 \pm{0.02}$  &  LT \\
57770.91 &  -  &  -  &  -  &  -  &  -  &  $20.33 \pm{0.04}$  &  $20.11 \pm{0.02}$  &  LT \\
57772.04 &  $21.98 \pm{0.01}$  &  $20.93 \pm{0.01}$  &  $20.25 \pm{0.01}$  &  $19.82 \pm{0.01}$  &  -  &  -  &  -  &  VLT \\
57772.91 &  -  &  -  &  -  &  -  &  -  &  $20.50 \pm{0.042}$  &  $20.22 \pm{0.02}$  &  LT \\
57778.88 &  -  &  -  &  -  &  -  &  -  &  $20.52 \pm{0.07}$  &  $20.27 \pm{0.03}$  &  LT \\
57779.22 &  $22.17 \pm{0.02}$  &  $21.23 \pm{0.02}$  &  $20.56 \pm{0.01}$  &  $20.08 \pm{0.01}$  &  -  &  -  &  -  &  VLT \\
57779.92 &  -  &  -  &  -  &  -  &  -  &  $20.73 \pm{0.03}$  &  $20.45 \pm{0.03}$  &  LT \\
57836.04&22.64$\pm$0.03 &-&21.24$\pm$0.01&-&-&-&-&VLT \\
   \hline
\end{tabular}
\label{table:logofphot}
\end{table*}


\begin{table*}
 \centering
 \caption{ Log of VLT spectroscopic observations. Note that all spectra were taken at an airmass of about 1.1 and sub-arcsecond seeing, and were acquired at parallctic angle}
 \begin{minipage}{350mm}
  \begin{tabular}{ccccccc}
  \hline
Epoch  &    Epoch    &       Phase              &   Setup    &      Range      & Filter & Exp. time\\
(UT)    &     (MJD)      &     (days\footnote{Rest frame.})     &     (Instr.+grism)         &  (\AA)   & & (s) \\
   \hline
2016 Dec. 22.05 &   57744.05   & 1.99  & FORS2+300V & 3300--6600 &$\cdots$ & 1800 \\
2016 Dec. 24.05 &   57746.05   & 3.73  & FORS2+300V & 3300--6600 &$\cdots$ & 1800 \\
2016 Dec. 26.10   & 57748.10   & 5.52  & FORS2+600B & 3300--6200 &$\cdots$ & 1800 \\    
2016 Dec. 29.07   & 57751.07  &  8.10  & FORS2+300V & 4000--8650 & gg435\footnote{Order separator.} & 1800 \\
2017 Jan. 01.26 &   57754.26   & 10.89 & FORS2+300V & 4000--8650 &gg435 & 1800 \\
2017 Jan. 05.06  &  57758.06   & 14.20 & FORS2+300V & 4000--8650 &gg435 & 1800 \\
2017 Jan. 07.08  &  57760.08   & 15.96 & FORS2+600B & 3300--6200 &$\cdots$      & 1800 \\
2017 Jan. 10.24  &  57763.24  &  18.71 & FORS2+300V & 3300--6600 &$\cdots$      & 1800 \\
2017 Jan. 14.15  &  57767.15  &  22.12 & FORS2+300V & 4000--8650 &gg435 & 1800 \\
2017 Jan. 26.24  &57779.24     & 32.65 & FORS2+300V & 4000--8650 &gg435 & 1800 \\
2017 Mar. 23.43  &   57835.43 & 81.69  & FORS2+300V & 4000--8650 &gg435 & 2400\\
2017 Oct. 18.27 & 58044.23 & 263.58 & X-Shooter     & 3000--22,000 &$\cdots$      & 4200  \\
2017 Oct. 23.27 & 58049.23 & 267.93 & X-Shooter     & 3000--22,000 &$\cdots$      & 4200 \\
   \hline
\end{tabular}
\label{table:logofspectra}
\end{minipage}
\end{table*}

\subsection{Optical photometry}

Photometric instrumental magnitudes were calculated using aperture photometry through a custom \textsc{python} pipeline utilising the \textsc{iraf} \textsc{daophot} package. FORS2 photometry was calibrated against secondary photometric standards  in the field of view of the target.  The local standards have been chosen among isolated stars in regions of constant background, and far away from bright and saturated objects. Their point-spread-function magnitudes were obtained with the same photometric zero point used to calibrate the magnitudes of stars in the Landolt 
fields SA95 and SA98 observed (respectively) before and after  SN\,2016jca on the nights of December 22 and 29. 
The LT photometry was calibrated to stars from the American Association of Variable Star Observers Photometric All-Sky Survey (APASS) Data Release 9. The TNG data reduction, including de-biasing and flat-fielding, was carried out following standard procedures: the cross-calibrated magnitudes were obtained using aperture photometry with respect to the secondary standard stars reported in Table \ref{table:std} of Appendix A.

\begin{figure*}
\centering
\includegraphics[scale=0.7]{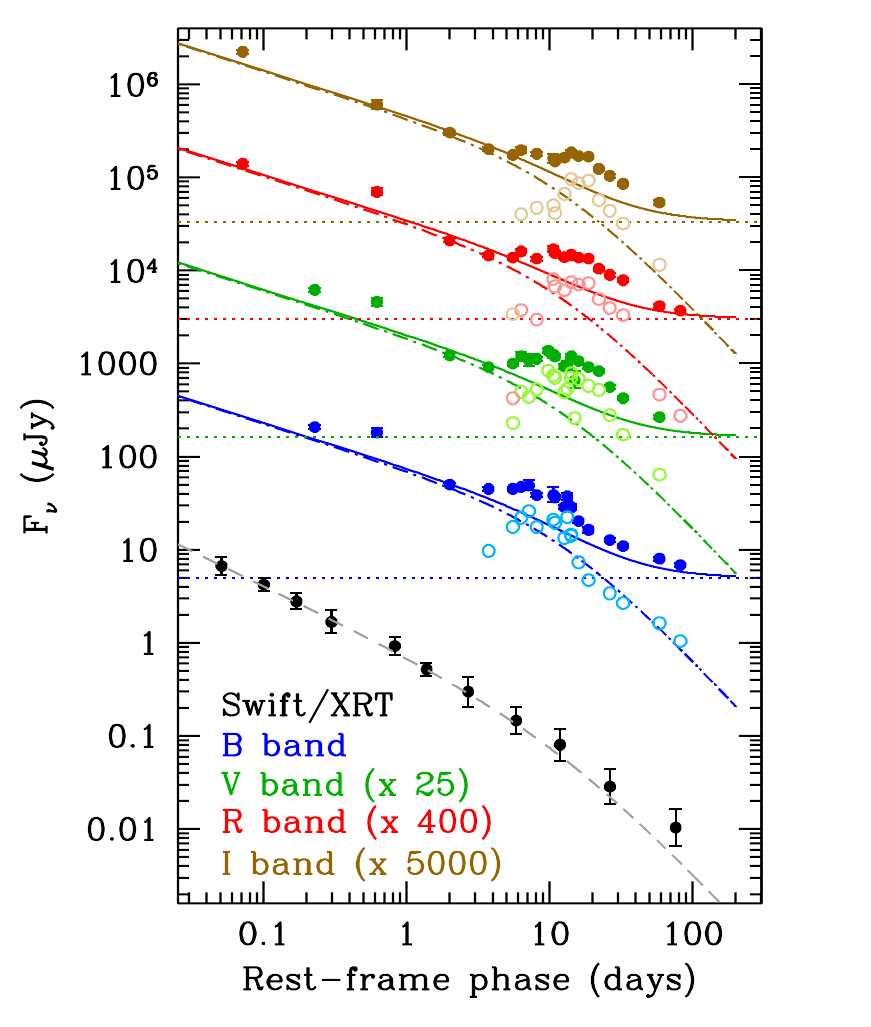}
\caption{Multiwavelength light curves of the GRB\,161219B counterpart.  The optical light curves in $BVRI$ filters (filled circles), constructed based on our and literature data, are K-corrected and deabsorbed for Galactic extinction.   The dotted horizontal line in each panel represents the host-galaxy contribution in that band; the dot-dashed curves represent the afterglow modelled with a steepening power law (see text).  The solid curve is the sum of the afterglow and host-galaxy components.   The subtraction of these two components from the observed points corresponds to the SN light curves in  $BVRI$ bands (light blue, light green, pink, and beige open circles, respectively).   The errors on the SN points are not reported for clarity and range from 10-15\% around maximum to 50\% at early epochs, when the contribution of the afterglow is larger.  The black filled circles  are  the {\it Swift}/XRT light curve  at 1\,keV;  the best-fitting afterglow model is reported as a gray dashed curve.}
\label{fig:decompLC} 
\end{figure*}

To the photometry  we applied a Galactic extinction correction using $E(B-V)_{Gal} = 0.028$\,mag \citep{Schlafly11} and the extinction curve of \citet{Cardelli89},  and as well a K-correction determined from our observed spectra. The resulting  $BVRI$ light curves are reported in Figure \ref{fig:decompLC}.  The detection of the GRB counterpart with all UVOT filters  and the multiwavelength spectral shape (see next section) at early epochs suggest that  intrinsic absorption due to dust in the host  galaxy is negligible, so we have not attempted to evaluate  a correction for such an effect.


\subsection{Optical spectroscopy}

Low-resolution spectra were acquired at the ESO Very Large Telescope (VLT) equipped with the FORS2 spectrograph using a variety of setups during the first 3 months after GRB detection (in the photospheric phase). During some of these observations the 300V grism was used and  also an order-separating filter was used, owing to severe contamination longward of 6000\,\AA\  by second-order light. These data are complemented with a VLT (plus X-Shooter) spectrum obtained on 2017 Oct. 18 and 23 (see Table \ref{table:logofspectra}).

The two-dimensional FORS2 spectral images  were flat-fielded and de-biased, and the one-dimensional spectra were optimally extracted \citep{Horne86} and reduced following standard procedures within  \textsc{iraf}. They were then linearised  and calibrated with respect to catalogued spectrophotometric standards, and corrected for atmospheric extinction. The flux calibration was refined by comparison with the  simultaneous broad-band photometry.  Telluric absorption lines and weak emission lines from the host galaxy were removed.

\begin{figure*}
\centering
\includegraphics[scale=0.7]{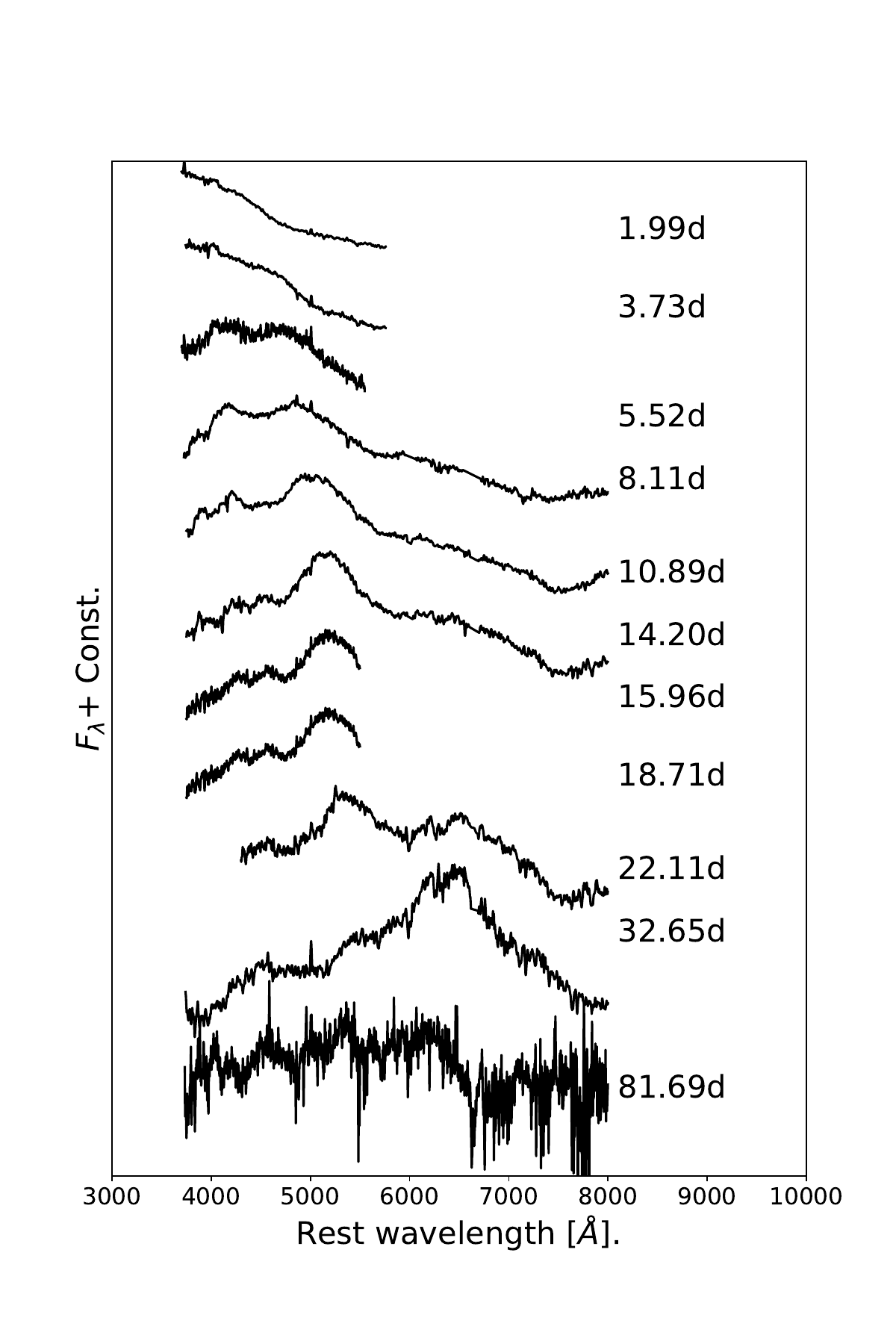}
\caption{A temporal series of spectra of  the transient. The time (in the rest frame) from explosion, $t$, is given for all 
spectra. The spectra are in the rest frame, corrected for foreground Galactic extinction and host-galaxy subtracted. This first spectrum appears to have no contribution from the SN component.}
\label{fig:spec}
\end{figure*}

Figure \ref{fig:spec} shows the FORS2 spectral sequence after correcting for  Galactic reddening and the host-galaxy contribution under the point-like source.   Both the wavelength scale and epochs are reported in the host-galaxy rest frame. The  most prominent emission lines of the host galaxy have been removed. The afterglow dominates the early emission: at $t=1.99$\,d the spectrum shows the usual afterglow power-law flux distribution, and no SN features are detected. (The low-level undulations are likely to be artifacts of the observation and data-reduction process (\citealt{Filippenko82}). The spectrum obtained at $t=3.73$\,d  still has a strong afterglow contribution, but it begins to show undulations seen in other GRB-SNe.  At $t=5.52$\,d, however, the spectral features typical of previously observed GRB-SNe are clearly seen, and by day 14.20 the observations look like a normal broad-lined SN~Ic.   The slope of the optical spectrum at $\sim2$\,d, when the SN and host-galaxy contributions are still negligible, is $\beta_{\rm opt} \approx 0.35$.

Simultaneous UV, optical, and near-infrared spectra ($\sim 3000$--21,000\,\AA) were taken with X-shooter using slit widths
of 1.0$^{\prime\prime}$,  0.9$^{\prime\prime}$, and 0.9$^{\prime\prime}$, for each arm, respectively \citep{DOdorico06}. We used a nodding throw along the slit (nodding lengths of 5$^{\prime\prime}$) to obtain better sky subtraction. The data were reduced using version xshoo/2.7.0b of the ESO X-shooter pipeline \citep{Modigliani10} with the calibration frames (biases, darks, arc lamps, and flat fields) taken during daytime. 
The spectra were extracted using standard \textsc{iraf} tasks. Spectrophotometric and telluric standard star exposures taken on the same night as SN\,2016jca observations were used to flux-calibrate the  spectra and to remove telluric features. 
The X-shooter spectra  were co-added to improve the signal-to-noise ratio. 

\begin{figure}
\centering
\includegraphics[scale=0.4]{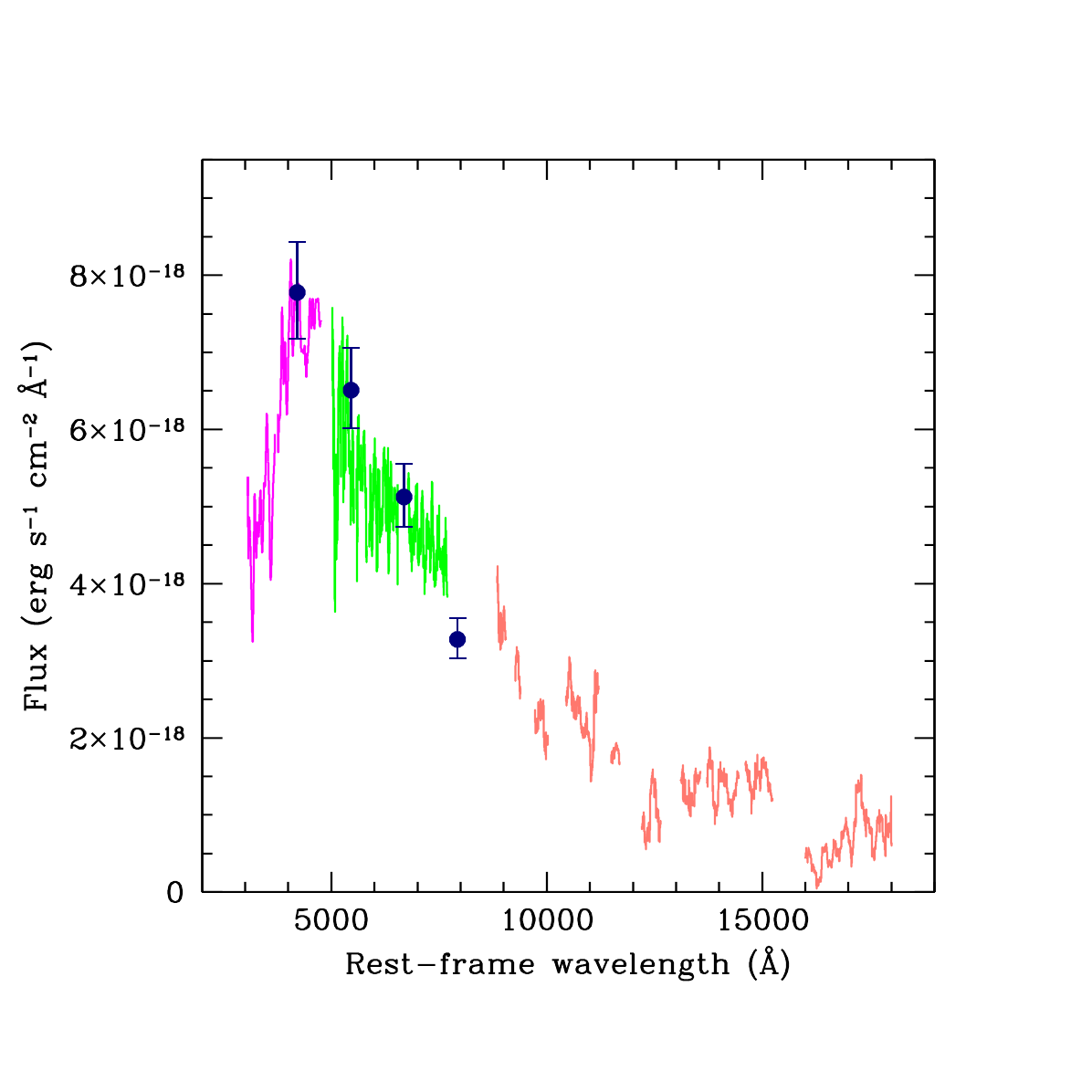}
\caption{VLT X-Shooter spectrum obtained from the average of observations taken on 2017 Oct. 18 and 23, with the near-UV (magenta), optical (green), and near-IR (orange) arms.  The spectrum is corrected for redshift and Galactic extinction and cleaned for artifacts and spurious features.  The dark-blue points indicate the extinction-corrected Pan-STARRS photometry of the host galaxy.  They are consistent with the spectral continuum, indicating that the latter is dominated by the host emission, with no evidence of nebular emission lines due to the SN. It is apparent from the spectrum that there is no sign of the SN at this epoch, +265\,d. }
\label{fig:latespec}
\end{figure}

The average spectrum, cleaned of artifacts and spurious emission features, is shown in Figure \ref{fig:latespec}. Since at this late epoch, +267\,d, the SN spectrum should be dominated by nebular  line emission, while the afterglow emission has completely faded, the spectral continuum is entirely due to the host galaxy, as confirmed by comparison with the archival Pan-STARRS photometry. Lack of detection of the nebular [O~I] $\lambda\lambda$6360, 6363 emission line (the strongest line typically detected at this late phase) leads to a 3$\sigma$ upper limit of $\sim  10^{-16}$\,erg\,s$^{-1}$\,cm$^{-2}$ on its intensity, corresponding to a line luminosity of $\sim 10^{40}$\,erg\,s$^{-1}$. Additional late-time spectroscopy was also conducted at the 10.4\,m GTC (+OSIRIS) on 2017 Jan. 17.

\subsection{X-ray and ultraviolet data}

\citet{Mingo16} provide a preliminary report on the {\it Swift}/XRT (0.3--10\,keV) observations of GRB\,161219B.  The   spectrum and light curve were downloaded from  the {\it Swift} archive (http://www.swift.ac.uk/xrt$\_$spectra/
and http://www.swift.ac.uk/xrt$\_$curves). Once corrected for Galactic ($N_{\rm H~I} = 3.06 \times 10^{20}$\,cm$^{-2}$) and intrinsic ($N_{\rm H~I} = (1.5 \pm 0.2) \times 10^{21}$\,cm$^{-2}$) neutral hydrogen absorption, the X-ray spectrum is described by a single power law $f(\nu) \propto \nu^{-\beta_{\rm X}}$,  $\beta_{\rm X} = 0.83 \pm 0.06$ 
(the uncertainty is 90\% confidence).  Using this spectral slope we have converted the X-ray fluxes into monochromatic fluxes at 1\,keV (Figure \ref{fig:decompLC}).

\begin{figure}
\centering
\includegraphics[scale=0.4]{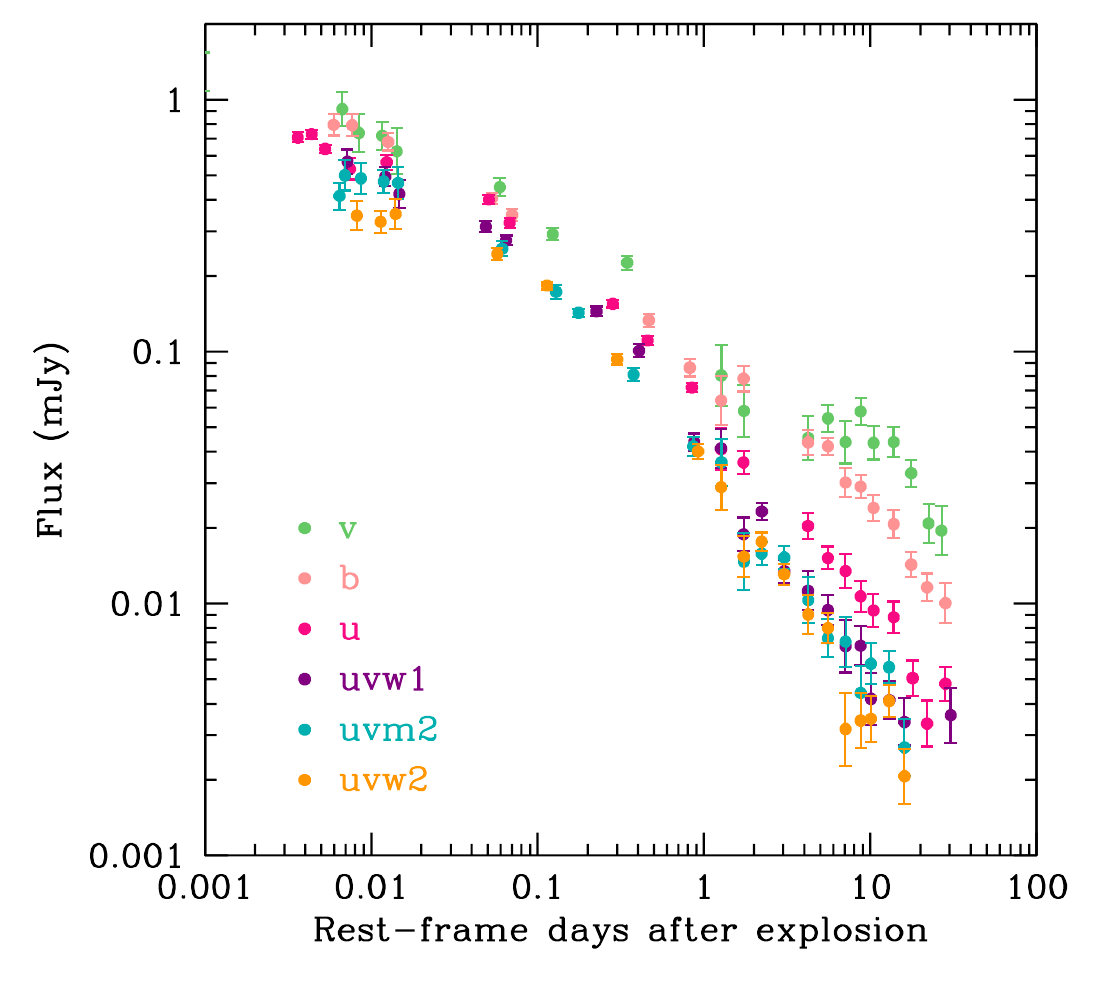}
\caption{UVOT light curves in six filters, corrected for Galactic extinction}
\label{fig:UVOTLCS}
\end{figure}

The {\it Swift}/UVOT began observing the field of GRB\,161219B 92\,s after the {\it Swift}/BAT trigger.  Observations were taken in both image and event modes. The afterglow
was detected in all seven UVOT filters.   Before extracting count rates 
from the event lists, the astrometry was refined following the methodology of 
\citet{Oates09}. The source counts were extracted initially using a source region of $5^{\prime\prime}$
radius. When the count rate dropped to below 0.5 counts per second, we used a source 
region of $3^{\prime\prime}$ radius. In order to be consistent with the UVOT calibration, these count 
rates were then corrected to  $5^{\prime\prime}$ using the curve of growth contained in the calibration files. 
As the afterglow is situated at the edge of the halo of a bright star, we used several small circular
regions at a similar distance around the halo in order to extract the background counts. The
count rates were obtained from the event and image lists using the {\it Swift} tools
{\tt uvotevtlc} and {\tt uvotsource}, respectively. They were converted to magnitudes
using the UVOT photometric zero points \citep{Poole08,Breeveld11}. The UVOT data, dereddened using the same procedure followed for the optical data, are shown in Figure \ref{fig:UVOTLCS}. The analysis pipeline used software HEADAS 6.19 and UVOT calibration
20150717. To improve the signal-to-noise ratio, the count rates in each filter were binned using
$\Delta t/t = 0.2$; this effectively bins only the late-time exposures.

The settling image is generally excluded as it may be affected by changes in the cathode voltage
during the first few seconds. We compared the magnitudes of several stars in the settling image
with later images and do not find a systematic difference; we therefore include this exposure
in our analysis.

\subsection{Decomposition of the multiwavelength light curves and spectra}

In order to study the SN component of the GRB optical counterpart, we must first subtract the contribution of the host galaxy and afterglow from the light curves and spectra of SN\,2016jca.
The host galaxy is clearly detected and resolved in both the VLT images and 
a {\it Hubble Space Telescope (HST)} image taken with WFC3/UVIS and the open LP filter (about 2000--10,000\,\AA); see \citet{Cano17}. 
We have identified on archival Pan-STARRS images of the host the precise position of the GRB-SN  and have measured the flux of the host within a circle of $1''$ radius.  
The $griz$ apparent magnitudes are $g = 23.0 \pm 0.1$, $r = 22.6 \pm 0.1$,  $i = 22.4 \pm 0.1$, and $z = 22.5  \pm 0.1$. 
The resulting colours are typical of  a modestly absorbed star-forming galaxy \citep{Kinney96}.  We have converted these magnitudes to the Bessel $BVRI$ system for subtraction from the optical counterpart.

While the initial temporal decline of the GRB optical counterpart can be attributed to a monotonically decreasing afterglow formed in a  blast wave,  starting around day 2--3  all light curves  show a rebrightening. We interpret this as SN emission, as clearly confirmed by the spectra that exhibit the typical stripped-envelope SN signature, especially at late times when the afterglow negligibly contributes to the total flux. 

\begin{figure}
\centering
\includegraphics[scale=0.4]{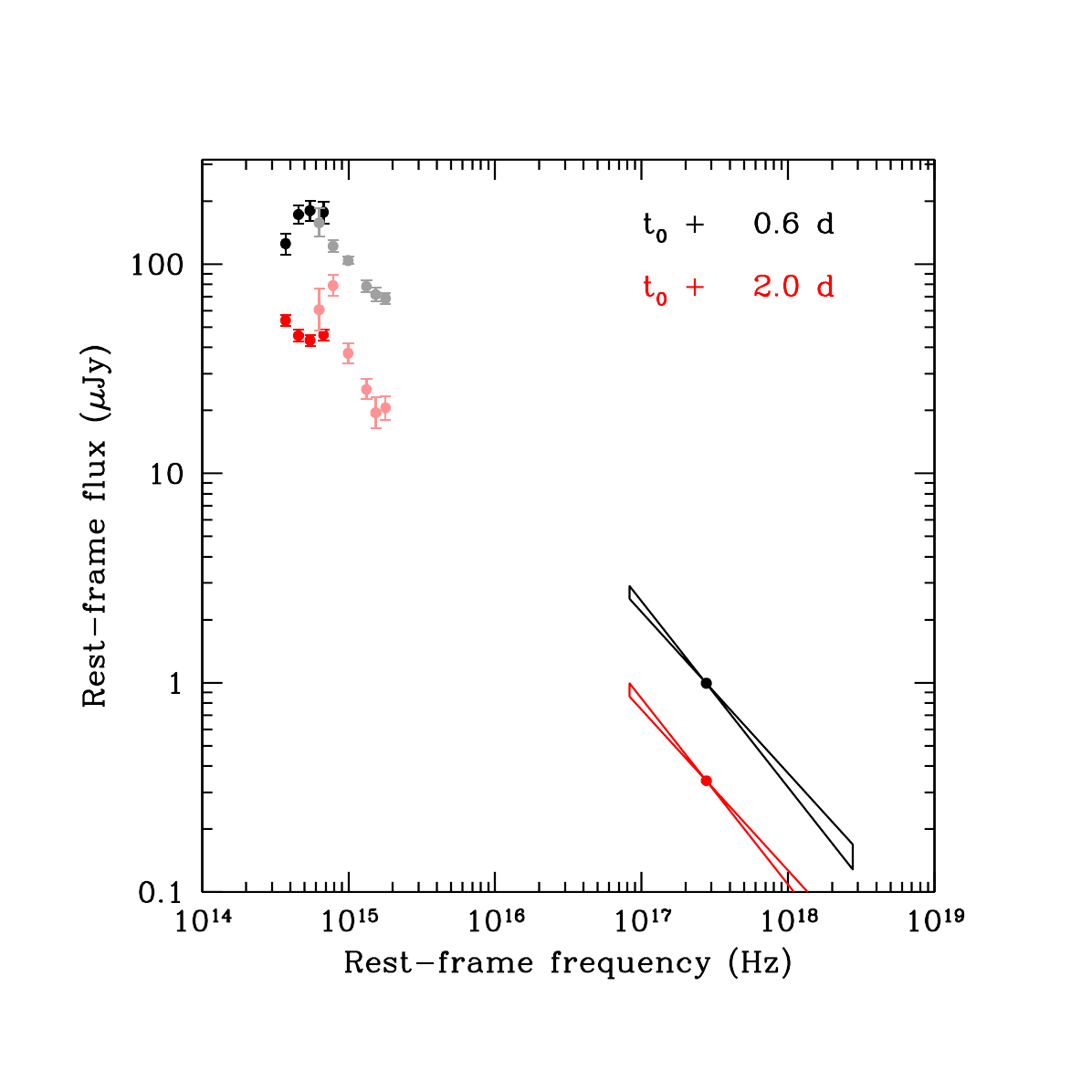}
\caption{SEDs of the GRB counterpart at 0.62 (black) and 2 (red) rest-frame days after explosion, corrected for Galactic and (for X-rays only) intrinsic extinction.  The optical data at the first epoch are from the literature, while at the second epoch they represent our VLT FORS2 photometry.   The simultaneous UVOT data are also reported in lighter shades (gray for $t_0 + 0.62$\,d and pink for $t_0 + 2$\,d).  The X-ray data are power-law spectral fits with index $\beta_{\rm X} = 0.83 \pm 0.06$. The host galaxy and SN flux are negligible with respect to the optical afterglow at these epochs and thus are not subtracted from the optical data. The optical magnitudes were converted to fluxes using the zeropoints from \citep{Fukugita95}.}
\label{fig:SED}
\end{figure}

The afterglow contribution was estimated  as follows.
We first constructed optical-to-X-ray spectral energy distributions (SEDs) at 0.62 and
2 rest-frame days. This corresponds to the first epoch that $BVRI$ photometry was reported and to the epoch of our 
first VLT/FORS spectrum, respectively. The SEDs are plotted in
Figure \ref{fig:SED}. The data are corrected for Galactic extinction, and in the X-ray band also for intrinsic extinction.  At these early epochs, the host-galaxy contribution is negligible. The SEDs and the spectral indices  estimated above in the individual domains ($\beta_{\rm opt} \approx 0.35$, $\beta_{\rm X} = 0.83 \pm 0.06$)  show that the optical broad-band spectrum has a significantly flatter slope than the X-ray spectrum and suggests that a cooling frequency, $\nu_c$
\citep{Sari98}, is between the optical and X-ray bands at these two epochs ($\sim 10^{15}$\,Hz). If the X-rays are produced by synchrotron
radiation, this implies that the electron energy distribution, $dN/d\gamma \propto \gamma^{-p}$, must have a 
slope $p = 2 \times \beta_{\rm X} = 1.66 \pm 0.12$ 
\citep{Sari98,Zhang04}.  
This corresponds to an optical spectral index $\beta_{\rm opt} = (p - 1)/2 = 0.36$,
in agreement with the spectral slope measured
in the $BVRI$ wavelength range.

Although the X-ray light curve  shows a deviation from a power law in the form of temporal steepening (see Figure \ref{fig:decompLC}),  it can be formally fitted with a single power law, $t^{-\alpha_{\rm X}}$, 
with $\alpha_{\rm X} = 0.88 \pm 0.05$.  According to standard fireball theory, the optical light curve should then decay with 
$\alpha_{\rm opt} \approx 0.63$.   However, this not only overpredicts the flux at epochs later than $\sim 20$\,d, but the residuals prior to day $\sim 20$ can only be accounted for by a SN that is a factor of 10--30 less luminous than SN\,1998bw. This is inconsistent with the spectra of SN\,2016jca that are approximately similar, at all epochs from few days to 3 months after explosion, to those of SN\,1998bw at comparable phases, implying that the temperature conditions (and therefore the luminosities) in the two SNe must be comparable; see Figure \ref{fig:compbw}.  Hence, the temporal modelling of the  multiwavelength afterglow must  account for some steepening, that we ascribe  to an achromatic  break in the  transition from a spherical expansion of the plasma to a jet geometry \citep{Sari99,Rhoads99}.

For $\nu > \nu_c$ and $p = 1.66$,  the standard fireball theory  of a relativistic shock propagating spherically in a constant-density circumstellar medium prescribes  that $\alpha = (3p + 10)/16 = 0.95$ 
 (``flat electron spectrum'' scenario, with $p < 2$; \citealt{Dai01}; \citealt{Zhang04}). 
However, this is  significantly steeper than our early measured X-ray decay rate.  On the other
hand, by applying  the ``classical'' steep electron  spectrum  scenario valid for $p > 2$, 
we predict $\alpha = (3p - 2)/4 = 0.75$, which is more consistent with observations.  

Therefore, we have a contradiction whereby, although the X-ray spectrum constrains $p$ to be less than 2, the X-ray 
light curve is consistent with fireball theory only for a scenario where $p > 2$.  In order to solve this discrepancy, we conclude that either  a standard impulsive blast-wave model is inapplicable in this case, or that the shallow decay is  due to a continuous, rather than impulsive, energy injection from the central engine  to the blast wave, which slows down the decay.
Since  fireball theory appears to be  a good description of the multiwavelength afterglow (see below), we assume that the latter hypothesis is more likely and postulate that the shock undergoes a re-energisation  episode whereby, starting  $\sim 0.1$\,d after the explosion (i.e., before the epoch  of the first SED; see Figure \ref{fig:SED}), energy in the form $L(t) \propto t^{-q}$  is injected in  the fireball.

In a spherical geometry, uniform-density medium and flat electron
spectrum ($p < 2$) scenario, the expected time-decay index is $\alpha  = [(2 + q)p + 18q - 12]/16$  below the cooling frequency
(i.e., the optical spectral range; see full derivation in Appendix A). Note that,   for $p = 1.66$  and 
$\alpha_{\rm opt} \approx 0.5$, we have $q \approx 0.85$.  Since $q < 1$, the blast-wave energy increases 
in time and it reduces the decay rates, as observed. Above the cooling frequency (X-ray band), 
the above $p$ and $q$ parameters imply a decrease rate of $t^{-0.8}$, consistent with the observed 
early-time decay of the X-ray light curve.  At late epochs, the predicted time-decay index in optical and X-rays is $\alpha_2 = p = 1.66.$

Then we determined the lower and upper boundaries of the achromatic break time of the light-curve power-law decay.    We adopted the temporal double power-law model of Israel et al. (1999), and studied the X-ray light curve, that does not contain a significant SN component and is exclusively due to synchrotron emission (any host galaxy component must also be negligible at the observed X-ray flux levels).  

After fixing the electron slope to $p = 1.66$,  we stepped the break time from 1 to 30\,d, with a 1\,d increase at each iteration.   For each value of the break time we explored 100 values of the flux normalisation,   stepped in a range that was iteratively adjusted until a $\chi^2$ minimum was found.    The formal best-fit break time is 26\,d,  with a lower boundary of 13\,d, determined following \citet{Avni76}.   However,  for  a break time of 26\,d or larger,  the luminosity of the SN resulting from the afterglow subtraction (see below)  is  incompatible with the temperature that is necessary to describe the spectra.  
For example, in spectral models of the SN,  for these late break times, to  keep the temperature and ionisation state constant in the models  the photospheric velocity would have to increase from day 3.73 to 5.52, which is unphysical. For a break time of 13\,d the model is compatible with the latest X-ray point lower boundary, and also returns an acceptable SN luminosity after subtraction.  Our best estimate of the break time is $t_{b} = 13 \pm 2$\,d.    An afterglow model was constructed based on the above parameters ($p = 1.66$,  $t_b = 13$\,d,  cooling frequency at 0.6\,d after explosion $\nu_C = 1.6 \times 10^{15}$\,Hz, flux at $\nu_C$ during $t_b$  $f_\nu =  6.2$\,$\mu$Jy) and on fireball theory.

If the achromatic time break is  related to the presence of a homogeneous jet where the plasma expands,  we estimate its
full opening angle to be $\theta = 42^\circ \pm 3^\circ$ \citep{Sari99},  assuming a
medium density of 1\,cm$^{-3}$ and the observed isotropic energy of $1.6 \times 10^{50}$\,erg \citep{Frederiks17}. 

We stress that the re-energisation of the blast wave, postulated to slow down the decay rates of the observed light curves, does not add any degree of freedom to the $\chi^2$ evaluation, its power-law index $q$ being determined univocally by the value of $p$ and being physically acceptable as long as it is $< 1$.  

\begin{figure}
\centering
\includegraphics[scale=0.4]{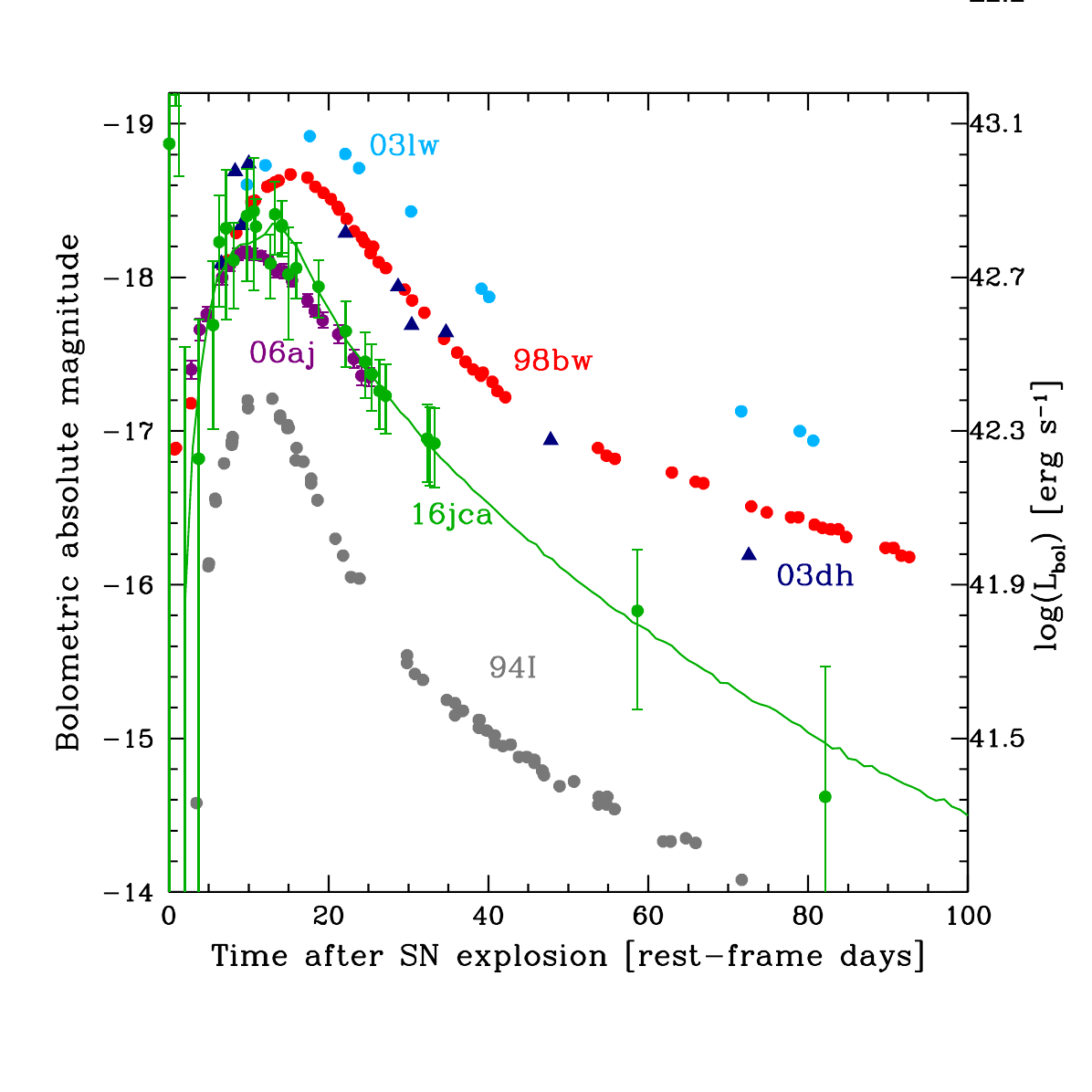}
\caption{Pseudo-bolometric light curve (3000--10,000\,\AA) of SN\,2016jca after the subtraction of the host galaxy and afterglow components, compared with bolometric light curves of previous GRB-SNe\,1998bw, 2003dh, 2003lw, and X-ray flash SN\,2006aj \citep{Galama98,Deng05,Malesani04,Pian06}, and the ``normal'' Type Ic SN\,1994I (not associated with a GRB) \citep{Richmond96}.   Note the similarity with the light curve of SN\,2006aj.  To avoid confusion, the error bars are shown only for SN\,2016jca.   The uncertainties of SN\,1998bw are smaller than the markers, SN\,2006aj are $\sim 10$\%, and SN\,2003dh and SN\,2003lw are comparable to those of SN\,2016jca. The first points, as in SN\,2003dh, are heavily contaminated by the afterglow. The solid green line is the model light curve produced with the density and abundance structure used for the spectra modelling.}
\label{fig:BolLC}
\end{figure}

By subtracting from the observed $BVRI$ light curves the constant host-galaxy flux and the afterglow component in each band as derived from our multiband fit using a break time of 13\,d, we obtained the $BVRI$ light curves of SN\,2016jca. These were used to construct, following a procedure similar to that adopted by \citet{Mazzali17},  a bolometric light curve in the range 3000--10,000\,\AA, which is plotted in Figure \ref{fig:BolLC}.

Similarly, we subtracted the host and the afterglow $BVRI$ SEDs from every dereddened and deredshifted 
spectrum of SN\,2016jca and used these decomposed spectra for comparison with SN models (see Section 4).

\section{The supernova component}

The peak of the bolometric light curve of SN\,2016jca (see Figure \ref{fig:BolLC})
occurred at $10 \pm 2$\,d after the GRB, 
at $-18.2 \pm 0.1$\,mag.  
A \Nifs\ mass of $\sim 0.27 \pm 0.05$\,\Msun\ is required to power the luminosity of the light curve if it is due to radioactivity, as is seen from 
the model in Figure \ref{fig:BolLC}.
 This \Nifs\ mass is similar to those of most other well-observed GRB-SNe.
 However, the rise time is 
shorter than that of other GRB-SNe by \ab2\,d, and significantly shorter than the 
prototypical GRB-SN, SN\,1998bw. 
This may indicate that the ejecta mass of SN\,2016jca is smaller 
than SN\,1998bw,  or it may imply that the viewing
angle to the SN symmetry axis is different, or that there is a 
larger abundance of \Nifs\ located further out in the ejecta in 
SN\,2016jca, or a combination of all three. 

\begin{figure}
 \centering
 \includegraphics[scale=0.3]{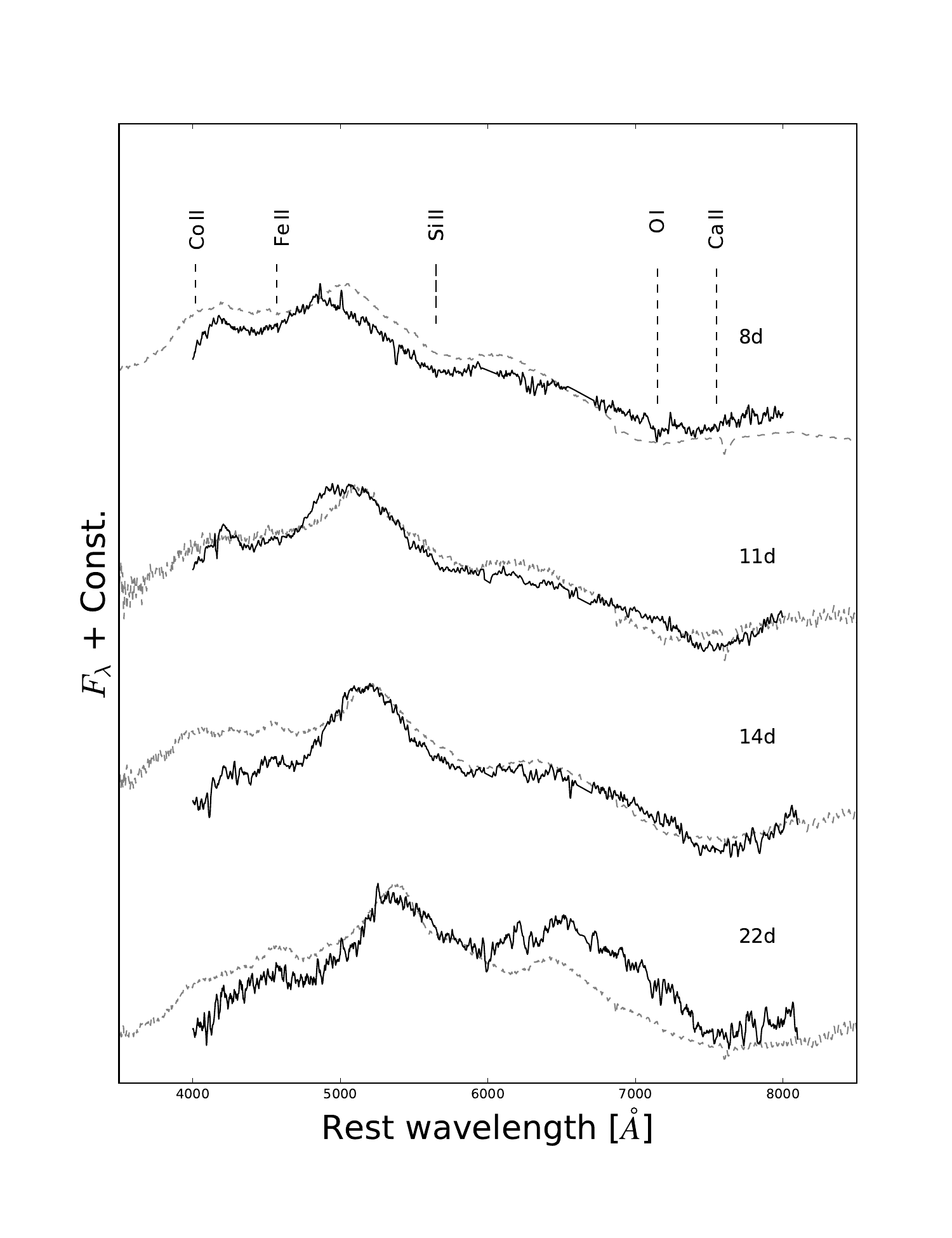}
 \caption{Spectra of SN\,2016jca (black) and SN\,1998bw (grey dotted)
 \citep{Iwamoto98} at $t = 8$, 11, 14, and 22 rest-frame days after explosion. 
 The spectra are in the rest frame, corrected for Galactic extinction 
 ($E(B-V)_{\rm Gal}$ = 0.028\,mag and 0.052\,mag, respectively) and host-galaxy subtracted.}
 \label{fig:compbw}
\end{figure}

Spectroscopically, SN\,2016jca has the standard broad absorption features attributed to
GRB-SNe. These are \FeII\ absorption at \ab4200\,\AA, \SiII\ absorption at 5800\,\AA, 
and the \OI/\CaII\ feature at 7600\,\AA.
The broad absorption features in GRB-SNe are caused
by a larger line-forming region compared to standard SNe~Ic. 
This larger region requires there to be sufficient densities at high 
velocities well above the photosphere to produce opacity.  Therefore, SNe with 
broader features have shallower density profiles \citep{Mazzali17}.  In effect, 
the gradient of the outer density profile of the ejecta 
determines the breadth of the lines: shallower profiles produce broader lines.
Hence, SNe with high specific \KE\ have broader lines. 
Spectroscopically SN\,2016jca is similar to SN\,1998bw, but SN\,2016jca 
has more absorption in the blue (see Figure \ref{fig:compbw}). 
This increased absorption is likely to be due to a higher abundance of Fe-group 
elements in SN\,2016jca, and we test this hypothesis below.

\section{Spectral Models}

\subsection{The code}

Having determined the afterglow properties and isolated the SN component, we now turn to spectral
modelling to analyse SN\,2016jca.  We use a one-dimensional Monte Carlo radiative-transfer code, 
which follows the propagation of photon packets through a SN atmosphere, to produce synthetic spectra. 
The code is based on work presented by \citet{Mazzali93}, \citep{Lucy99}, and \citep{Mazzali00}.
This technique utilises the fact that as time passes, progressively deeper layers of the 
SN ejecta can be seen, since a SN can be assumed to be in homologous expansion starting $\sim10$\,s after explosion. 
We have $r=v_{\rm ph} t_{\rm exp}$, where $r$ is the distance form the centre of the explosion,  $v_{\rm ph}$ is the photospheric 
velocity, and $t_{\rm exp}$ is the time from explosion. The Schuster-Schwarzschild approximation is used, which assumes that 
the radiative energy is emitted at an inner boundary in the form of a black body. This is a sound approximation
for modelling a GRB-SN, as it yields good results, owing to the amount of material above the photosphere.
Furthermore, the approximation does not require in-depth knowledge about the radiation  transport below the photosphere.

The code has previously been successfully used for Type Ia SNe \citep[\eg][]{Ashall14,Ashall16,Ashall18} as well as core-collapse SNe \citep[\eg][]{Mazzali17,Prentice18}. A detailed explanation of the modelling procedure as well as the error analysis can be found in the companion paper \citep{Ashall19}.
In short, the purpose of the code is to produce optimally fitting synthetic spectra. This is done by varying input parameters,
such as the bolometric luminosity, $v_{\rm ph}$, and abundances, given a fixed input density profile. 
While  using a one-dimensional code to infer the structure of the event represents an inherent limitation for the characterisation of certain properties such as asymmetry, a higher-dimensional code requires that a very large set of assumptions (e.g., asphericity in density/velocity, orientation) must be made, such that it may be very difficult (if not impossible) to distinguish between different options. The approach using a one-dimensional code removes most of these assumptions and allows us to at least get a direct glimpse of one of the most likely reasons for the behaviour of the spectra.

The  mass we obtained for SN\,2016jca was lower than that for SN\,1998bw, as SN\,2016jca has a more rapidly evolving light curve. The ejecta mass and \KE\ of the 
density profile used were 6.5\,\Msun\ and $4 \times 10^{52}$\,erg (respectively), although the formal values we adopt for the analysis are  $6.5 \pm 1.5$\,\Msun\ with a range of \KE\ of  (3.0--5.0) $\times 10^{52}$\,erg. In the modeling procedure we used  
constant abundances as a function of velocity,  
and our most abundant elements 
are O(\ab70\%), Ne(\ab20\%), and C(\ab7\%), followed by Si(\ab1.5\%), S(\ab0.5\%), and \Nifs(\ab0.4\%), with the remaining 0.5\% consisting of Mg, Ca, Fe, and Ti+Cr.  

\begin{figure*}
\centering
\includegraphics[scale=0.3]{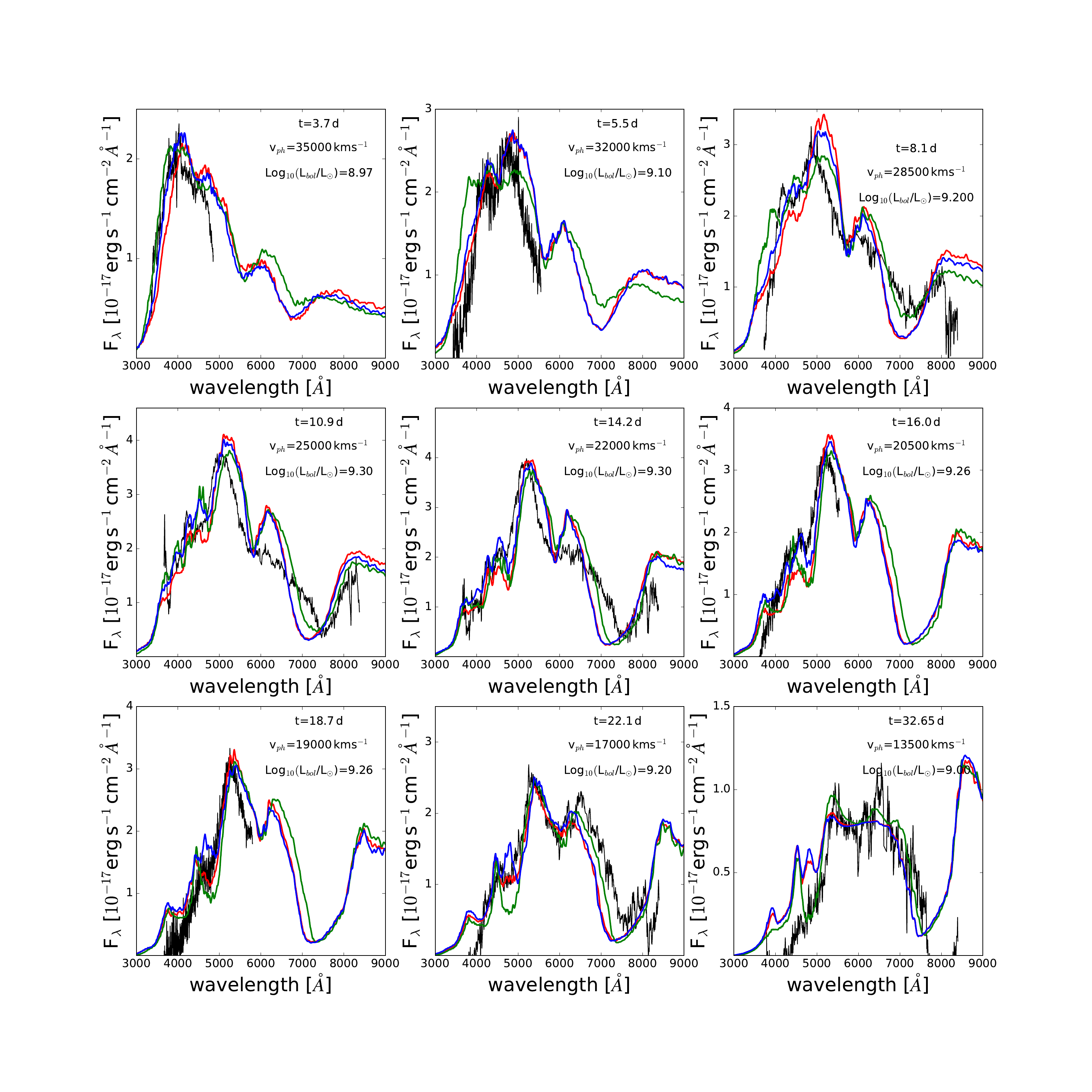}
\caption{Spectral models produced with \Nifs\ abundances. 
The best-fit blue model has a \Nifs\ abundance of \ab0.4\%, or a \Nifs\ mass fraction $X_{\rm{Ni}}$ of 0.004; note that this is the mass fraction relative to the total ejected mass. The red model has a \Nifs\ abundance which decreases as velocity decreases, and the green model has the opposite trend.}
\label{fig:stratNi}
\end{figure*}

However, having a constant mass fraction of \Nifs\ as a function of velocity is difficult to reconcile with an 
aspherical explosion, especially if the \Nifs\ is produced on the side of the GRB-jet. It would be expected that 
as the photosphere recedes, the jet, and the heavily synthesised region surrounding the jet, becomes a smaller overall fraction 
of the total observed region. Therefore, if the \Nifs\ was produced in the region surrounding the jet it could be 
expected that its abundance decreases as a function of decreasing velocity. 
Thus, we test models where the \Nifs\ abundance changes (both increasing and decreasing) as a function  of velocity.
Figure \ref{fig:stratNi} presents the models, as well as models with a flat \Nifs\ distribution. 
The (green) model where the \Nifs\ abundance increases as a function of decreasing velocity produces poor fits. 
At early times (5.5, 8.1, and 10.9\,d) there is not enough blocking by \NiII\ and \CoII\ lines, owing to their low abundance.
Conversely,  at later times (22.12\,d) there is too much absorption at $\sim4200$\,\AA, caused by the decay of \Nifs. 
The (blue) models with constant \Nifs\ abundance of 0.4\% produce good fits throughout, but 
the (red) models where the \Nifs\ abundance decreases as a function of decreasing velocity produce fits which are better or as good as the blue models.

\begin{figure}
\centering
\includegraphics[scale=0.6]{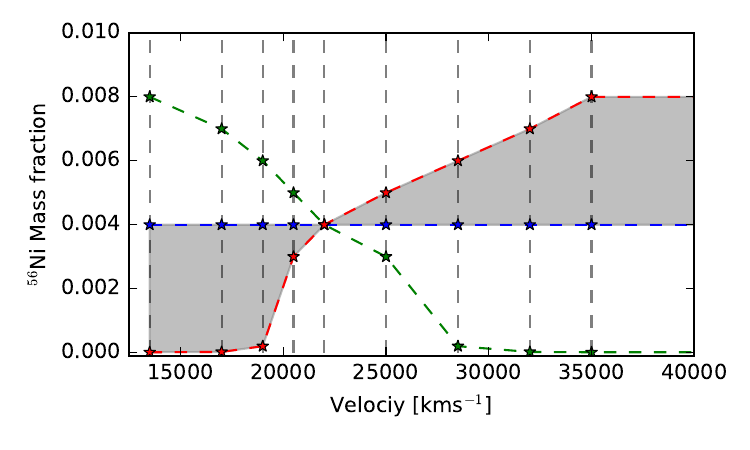}
\caption{ The \Nifs\ abundance distribution as a function of velocity for the three models presented in Figure \ref{fig:stratNi}. The values of \Nifs\ which produce good fits are highlighted in grey. The dashed vertical lines represent the photospheric velocity from the nine spectral models. Note that this is the mass fraction relative to the total ejected mass.}
\label{fig:stratNiab}
\end{figure}

\begin{figure}
\centering
\includegraphics[scale=0.3]{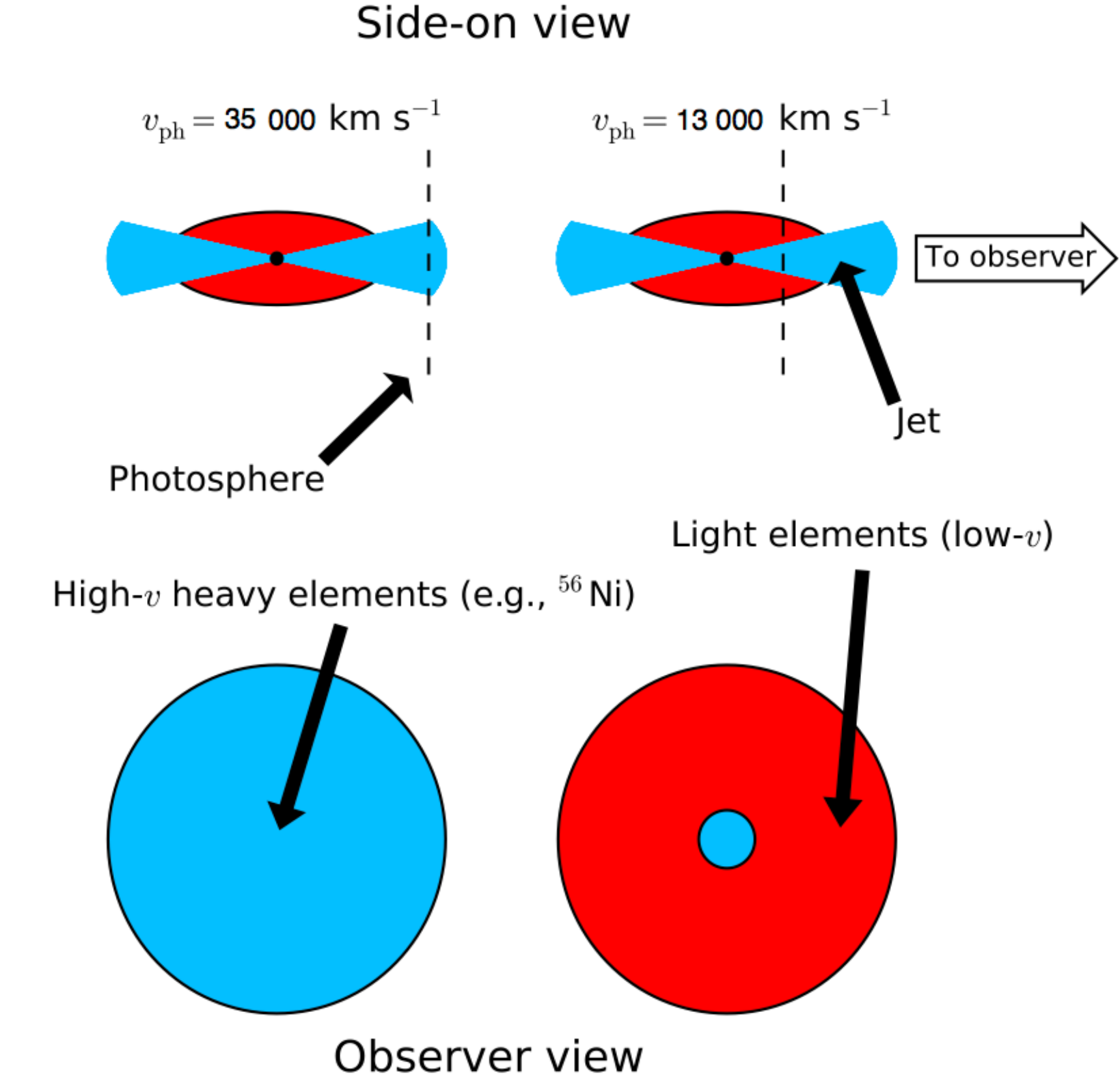}
\caption{A demonstration of the ejecta velocity distribution in an aspherical explosion with jet component and the dependence on photospheric velocity $v_\mathrm{ph}$ for spectral line velocities. All material to the left of the photosphere is optically thick to the observer, while material
to the right is optically thin; this is where line formation occurs.
(Left) At $v_\mathrm{ph} = 43,000$\,km\,s$^{-1}$, the photosphere forms inside the jet, and so only high-velocity material contributes to spectral line formation. (Right) At a lower $v_\mathrm{ph}$ the photosphere forms across both the high- and low-velocity material, providing a range of line velocities in the spectra. The figure is for illustrative purposes only. }
\label{fig:view}
\end{figure}

Therefore, we have a range of possible \Nifs\ values between the red and blue models which produce good fits. 
These are shown in  Figure \ref{fig:stratNiab}; the shaded region shows the uncertainty of the 
\Nifs\ abundance. 
There is evidence that the \Nifs\ abundance decreases as the velocity decreases. 
This could be can be explained if an aspherical explosion placed heavy elements in and 
around the region where the explosion is more energetic.
At the earliest epochs, the observed photosphere would  
consist of more metal-rich material at the highest velocities, 
and as the photosphere recedes the abundance of
lighter material on the side of the ejecta would increase; an example of this can be seen in Figure \ref{fig:view},
and the metal abundance decreases. 
This could be indirect evidence that the SN is aspherical both in shape and in elemental distribution.


\begin{figure*}
\centering
\includegraphics[scale=0.3]{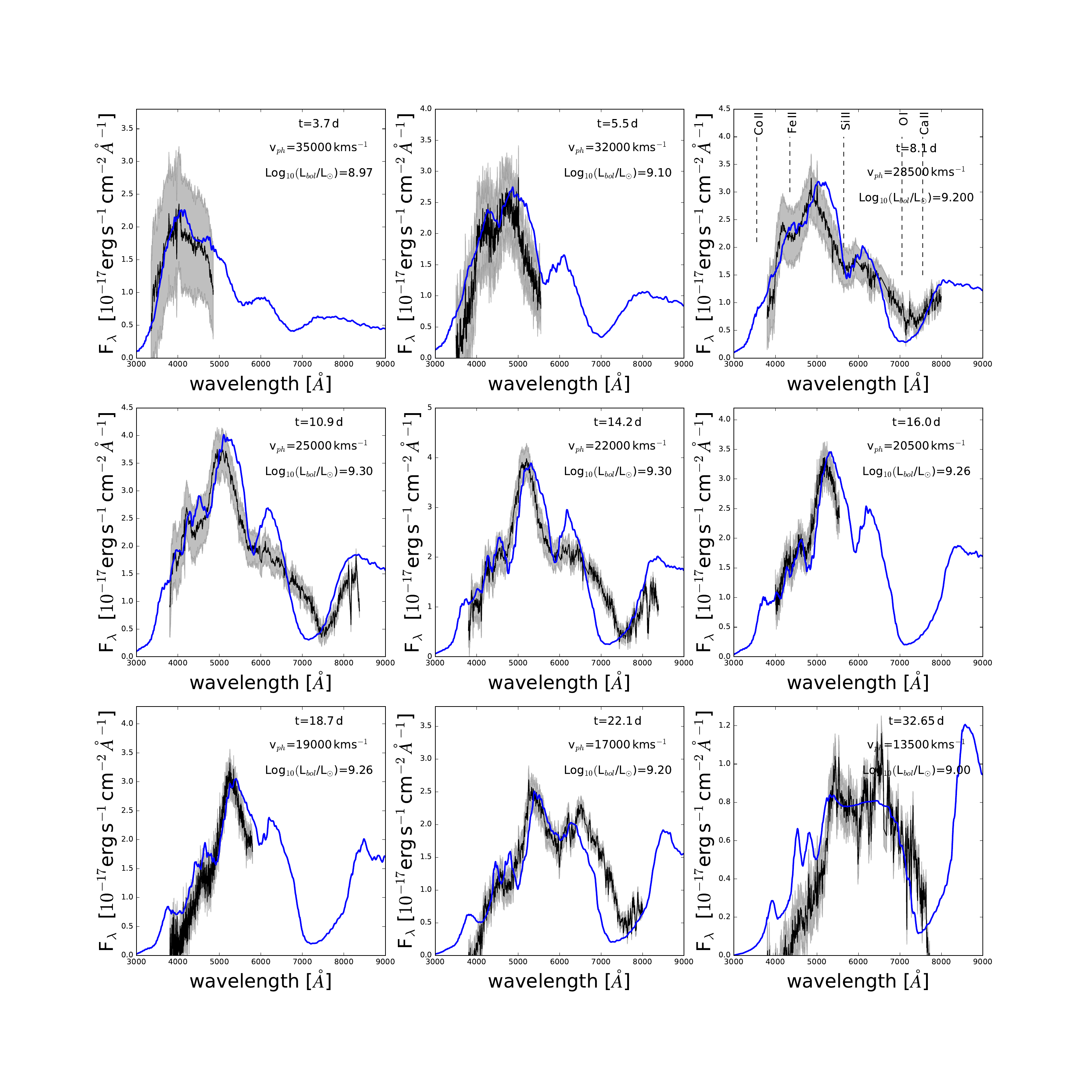}
\caption{VLT (+ FORS2) spectra of SN\,2016jca in the rest frame (black) and the best-fit models (blue) at nine epochs. These models have a decreasing \Nifs\ abundance as a function of decreasing velocity.  The spectra are corrected for Galactic extinction ($E(B-V)_{\rm Gal} = 0.028$\,mag) and smoothed with a 15\,\AA\ boxcar.  The host galaxy and afterglow components were subtracted.  The shaded grey region represents the uncertainty related to afterglow subtraction.}
\label{fig:models}
\end{figure*}

\subsection{Best models}

Figure \ref{fig:models} presents our optimised models; they were produced with 
the input parameters given in Table \ref{table:logofmodel}.
The earliest model at 3.7\,d is very blue, hot, and almost featureless. 
Absorption and reprocessing of blue flux is caused by blends of metal lines
including \NiII\ $\lambda\lambda$3465, 3471, 3513, 3576, 4067, \CoII\ $\lambda\lambda$3388, 3415, 3446, as well as \MgII\ $\lambda\lambda$2795, 2802 resonance lines and \CaII\ $\lambda\lambda$3934, 3969.
The model at this early epoch  has weak \SiII\ $\lambda\lambda$6347, 6371 absorption at 5300\,\AA, 
and \OI\ $\lambda\lambda$7772, 7774, 7775 and \CaII\ $\lambda\lambda$8498, 8542, 8662 absorption  at \ab6500\,\AA. At this epoch the Sobelov optical depths of the \CaII\ lines are a fifth of the corresponding \OI\ lines.

Conversely, $\sim 2$\,d later at 5.5\,d, the model spectrum contains stronger 
features typically associated with broad-line SNe~Ic. 
The blue region is still depressed in flux owing to 
line blanketing caused by blends of metal lines, and the 
absorption at 4200\,\AA\ is produced by \FeII $\lambda\lambda$5018, 5169 with 
some contribution from \SiII\ $\lambda$5056. 
The \CaII\ near-infrared triplet absorption is stronger than in the previous epoch,
and it occurs redward of the corresponding \OI\ absorption. This 
is unlike in SN\,1998bw, where the \CaII\ absorption is at shorter wavelengths and 
higher velocities than that of the \OI\ lines. The degree of blending in SN\,2016jca 
indicates that
the mass above 0.1c is significant, and our models show \CaII\ absorption up to
0.25c.

As time passes the SN ejecta expand and cool, but increase in luminosity. 
The peak in bolometric luminosity in the models is reached
between the 10.9\,d and 14.2\,d;
this coincides with the bolometric maximum derived 
from the light curve at \ab12\,d (see Figure \ref{fig:BolLC}).  
The models match the observations well at all epochs. 

Our models have 0.5\,\Msun\ of material above 0.1c (of which $2 \times 10^{-3}$\,\Msun\ is \Nifs) and
0.1\,\Msun\ of material (of which $4 \times 10^{-4}$\,\Msun\ is \Nifs) above 0.15c.
The presence of
this mass at such high velocities may represent a connection with the relativistic
outflow \citep{Piran2017}. These values should be reduced to take into account the
likely asphericity of the SN ejecta, with equatorial material carrying less \KE.
Regardless of this, the mass of \Nifs\ at high velocity
is much less than what is required to drive the light curve
($\sim 0.27\pm0.05$\,\Msun), and as there is $\sim 0.03$\,\Msun\ of \Nifs\ above 10,000\,\kms, we conclude that most of the \Nifs\ ($\sim 0.24$\,\Msun) must be
located at lower velocities. This \Nifs\ can efficiently deposit its decay
products (gamma-rays and positrons) at advanced epochs and power the spectrum
during the nebular phase ($\sim 200$\,d after explosion). Creating \Nifs\ at low
velocities poses a challenge to GRB-SN models. 

\begin{table}
 \centering
 \caption{Input parameters for the spectral models }
  \begin{tabular}{ccc}
  \hline
Epoch &   $v_{\rm ph}$   &  $L$   \\
(rest-frame days)& km\,s$^{-1}$& log($L$/$L_{\odot}$) \\
  \hline
  3.73& 35,000$\pm$7000&8.97$\pm$0.1 \\
  5.52& 32,000$\pm$6400&9.10$\pm$0.05 \\
 8.11& 28,500$\pm$2850&9.20$\pm$0.02 \\
10.89& 25,000$\pm$1250&9.30$\pm$0.02 \\
14.20& 22,000$\pm$1100&9.30$\pm$0.02\\
15.96& 20,500$\pm$1035&9.26$\pm$0.02 \\
18.71& 19,000$\pm$950&9.26$\pm$0.02 \\
22.12& 17,000$\pm$850&9.20$\pm$0.02 \\
32.65& 13,500$\pm$675&9.00$\pm$0.02 \\
  \hline
\end{tabular}
\label{table:logofmodel}
\end{table}

\subsection{Velocities}

\begin{figure}
\centering
\includegraphics[scale=0.4]{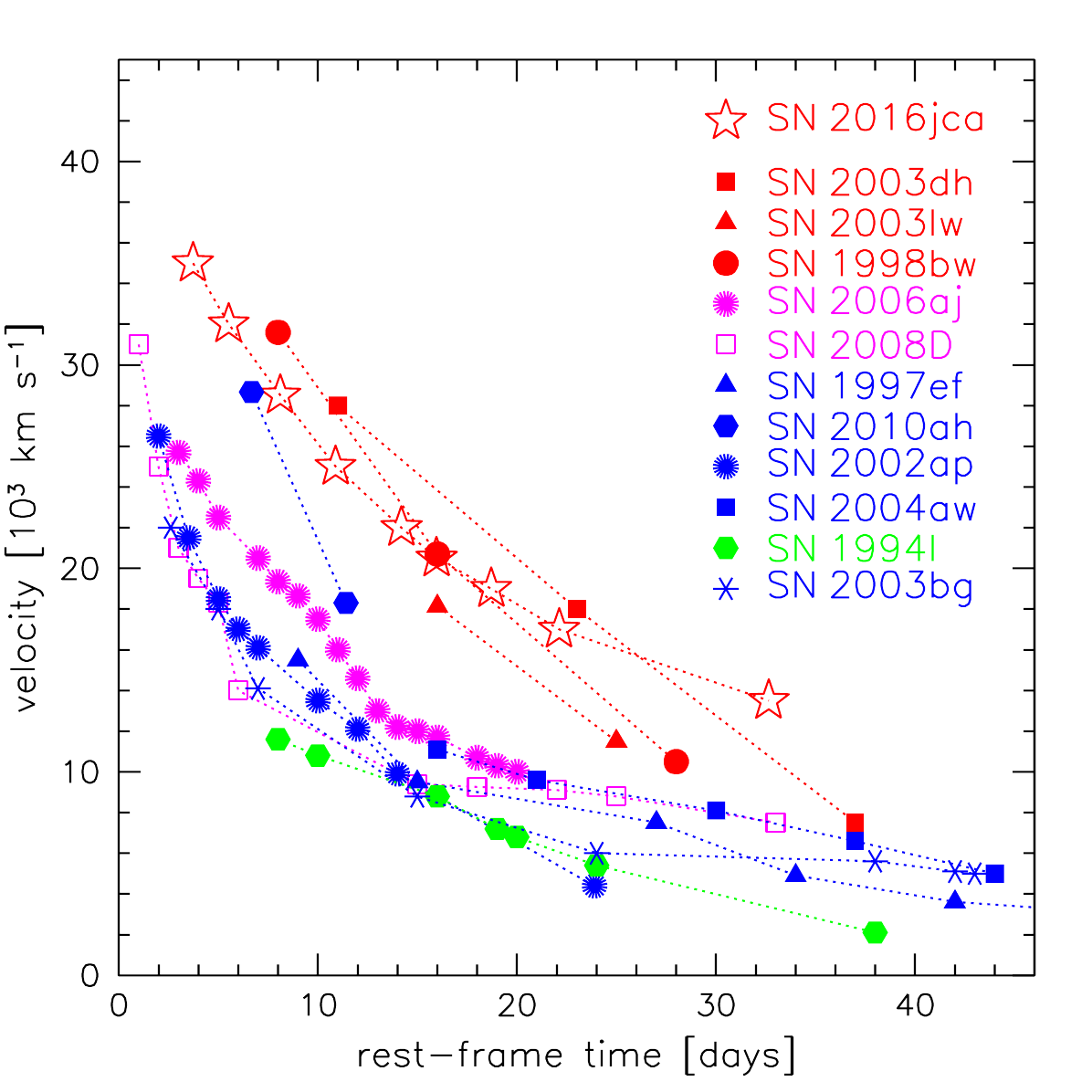}
\caption{Photospheric velocity, as determined from spectral models, as a function of time from explosion for 12 modelled SNe. 
The red points are  GRB-SNe, the pink points are SNe associated with X-ray flashes. The blue symbols
are high-\KE\ SNe with no detected accompanying high-energy event.  SN\,1994I is a normal SN~Ic.
 The typical uncertainty of $v_{\rm ph}$ is 
$\sim 20$\%.}
\label{fig:vel}
\end{figure}

Figure \ref{fig:vel} presents the photospheric velocity evolution as a function of time for 12 stripped-envelope SNe. Our comparison of photospheric velocities is based only upon quantities determined with the aid of a spectral model.
We do not consider velocities measured from the observed spectra, as these have several drawbacks: observed absorption lines form above the photosphere, and therefore indicate velocities higher than the photospheric velocity and which depend on the strength of the lines. 
Additionally, lines are usually blended, especially if broad, making it difficult or impossible to isolate the contribution of individual atomic species.

It can be seen in Figure \ref{fig:vel} that the velocity of the photosphere drops rapidly over the first 5\,d.  SN\,2016jca was observed as early as SN\,2002ap \citep{Mazzali02}; however,  it shows much higher velocities. The velocity of SN\,2016jca at 3.73\,d was $35,000 \pm 7000$\,\kms, and this is
the largest photospheric velocity recorded for a GRB-SN.
At later times, SN\,2016jca evolves like other GRB-SNe, which highlights the similarity among these events as well as their exceptionally high energies. With the exception of SN\,2010ah there is a clear difference in the 
velocity evolution of GRB-SNe and other striped envelope SNe (including those associated with X-ray flashes).
GRB-SNe have consistently higher photospheric velocities than standard events, and this trend is also seen with their \KE.

SN\,2010bh was previously  interpreted as the highest velocity GRB-SN ever discovered. However, in most observational papers velocities are obtained  by measuring the minima of a feature --- yet as we discussed above, using one transition as a proxy is extremely uncertain. 
\cite{Bufano2012} find that for SN\,2010bh, at 2.4\,d after explosion,
 there is \CaII\ absorption at \ab47,000\,\kms.
 If we follow the same approach, measuring the minimum of the feature of our synthetic spectra, 
 we obtain a a velocity of  \ab71,700\,\kms. Furthermore, 
 if we take a direct measurement from the observed spectra at 8.1\,d,
 the minimum of the O/Ca absorption feature is at \ab7200\,\AA, 
which corresponds to a velocity of 47,000\,\kms.
 This value is the same as that in \cite{Bufano2012} but measured 5\,d later, which indicates 
that the earlier velocities in SN\,2016jca must be larger. 
We note that \cite{Toy16} present a spectrum of SN\,2013dx (which has been afterglow subtracted) at 3.25\,d, but their spectra is too noisy for a velocity measurement, with their first measurement being at 9.3\,d.  
In this work we use radiative-transfer models to consistently calculate
the photospheric velocity and compare this to other models made with the same radiative-transfer code, which allows for consistency.  However, 
as shown above, if we use the same measuring techniques as in other papers, SN\,2016jca still has the highest velocity features.

\begin{table*}

  \caption{Properties of GRB-SNe closer than $z =0.3$. Properties are taken from \citep{Mazzali14} and references therein.}
  \begin{tabular}{ccccccccc}
  \hline
  GRB-SN   	 &  $z$   &  T90   &   $E_{\rm iso}$  & $\theta_{\rm op}$ &  $E_{\gamma}$  &   SN \KE\      &    $M(^{56}$Ni)&    $E_{\rm radio}$     \\ 
                 &        &   [s]  &  [$10^{50}$\,erg] &   [deg]   &   [$10^{50}$\,erg] & [$10^{50}$\,erg]&   [$M_\odot$]  &  [$10^{50}$\,erg]   \\
        1        &    2   &    3   &        4         &     5     &         6         &        7       &       8        &         9          \\
 \hline
980425  / 1998bw & 0.0085 &   30 & $0.010 \pm 0.002$ & 180        & $0.010 \pm 0.002$ & $500 \pm  50$ & $0.43 \pm 0.05$ & $\sim 0.2$          \\
030329  / 2003dh & 0.1685 &   23 & $  150 \pm 30$    & $6 \pm 2$  & $0.23 \pm  0.05$  & $400 \pm 100$ & $0.4 \pm 0.1$   & $2.5 \pm 0.8$      \\
031203  / 2003lw & 0.1055 &   40 & $  1.0 \pm 0.4$   & 180        & $1.0 \pm 0.4$     & $600 \pm 100$ & $0.6 \pm 0.1$   & $0.17 \pm 0.06$   \\
060218  / 2006aj & 0.0335 & 2000 & $ 0.53 \pm 0.03$  & 180        & $0.53 \pm 0.03$   & $20 \pm 6$ & $0.20 \pm 0.05$ & $0.020 \pm 0.006$\\
100316D / 2010bh & 0.059  & $>$1300 & $ 0.7 \pm 0.2$ & 180        & $0.7 \pm  0.2$    & $100 \pm  60$ & $0.12 \pm 0.02$ & $\sim 0.2$         \\ 
120422A / 2012bz & 0.283  &    5 & $ 2.4  \pm 0.8$   & $23 \pm 7$ & $0.05  \pm 0.02$  & $400 \pm 100$ & $0.3 \pm  0.1$  &  $\cdots$ \\
130427A / 2013cq & 0.3399 &  160 & $8100  \pm 800$   & $3 \pm 1$  & $4 \pm 1$     & $640 \pm  70$   & $0.4 \pm 0.1$ & $6 \pm 2$           \\
130702A / 2013dx & 0.145  &   59 & $ 6.5 \pm  1.0^a$ & $14 \pm 4$ & $0.05 \pm 0.02$   & $300 \pm  60$ & $0.3 \pm 0.1$   & $20 \pm 5$       \\
161219B/ 2016jca & 0.145  &7&  1.6   &   42$\pm$3      & 0.1 &   400$\pm$80       &    0.30$\pm$0.05                   &    $\cdots$             \\
\hline
\end{tabular}
\label{table:GRBSNprop}
\end{table*}


\begin{figure*}
\centering
\includegraphics[width=\textwidth]{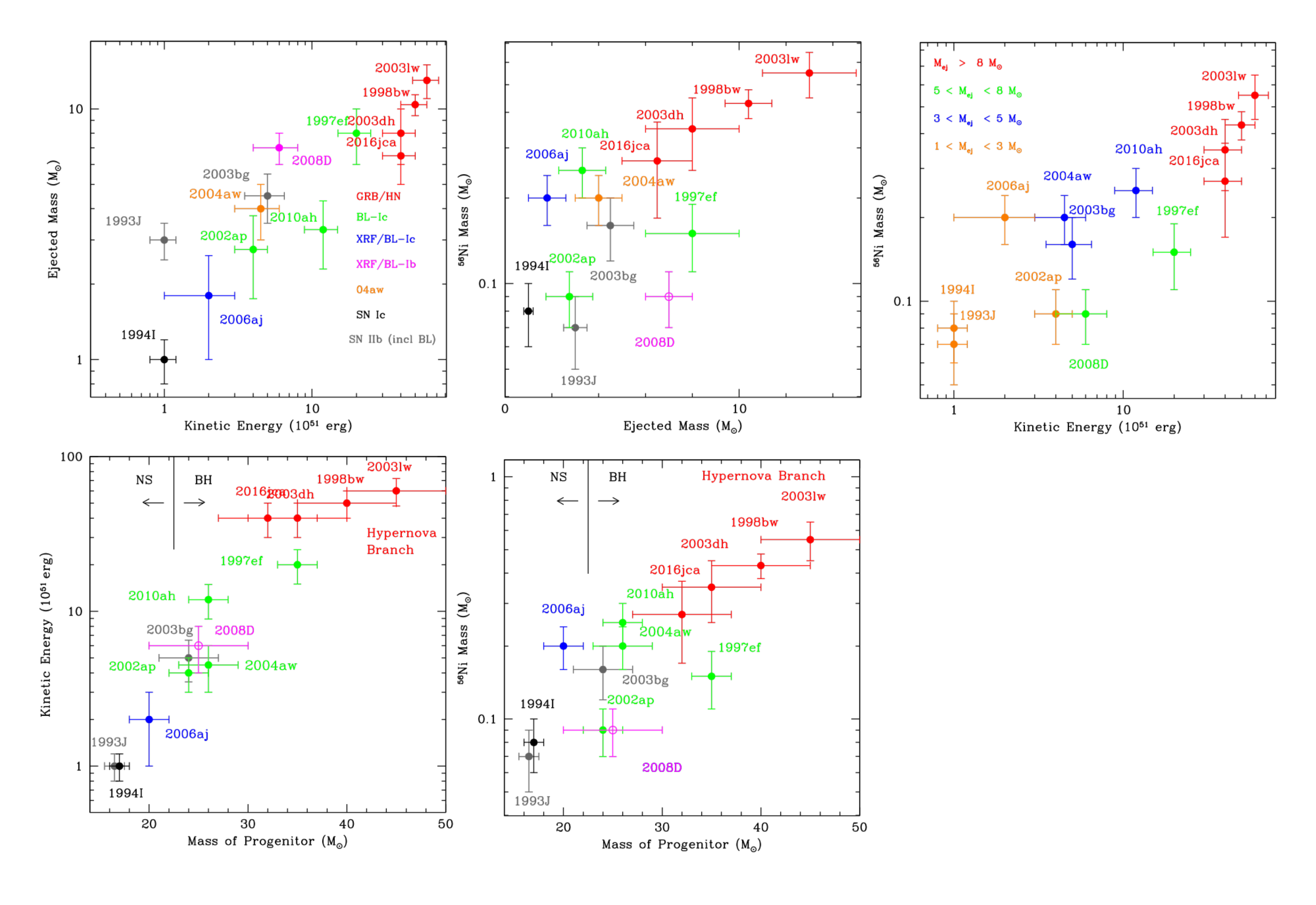}
\caption{ The location of SN 2016jca in terms of the stripped-envelope-SN
correlations from \citet{Mazzali17}. For the left and
middle panels, the red points are GRB-SNe; the blue point is SN2006aj, a
SN Ic associated with and X-ray flash; the pink point is SN2008D,  a SN
Ib associated with an X-ray flash; the black point is SN1994I, a normal
SN Ic; the orange point is SN2004aw, where a fast evolution of photospheric velocity (and therefore line width) was proposed \citet{Mazzali17};  the green points are broad-lined SNe Ic, and the
grey points are SNe IIb. For the right top panel the SNe are colored
based on ejecta mass.}
\label{fig:corelations}
\end{figure*}

\subsection{Model light curve}
To check our results and to verify the density profile and abundance structure we have used in our analysis, we have produced a model light curve.
We use a Monte Carlo light-curve code which was first presented by \cite{Cappellaro97} and expanded by \cite{Mazzali00}. Using the density profile and abundance structure from our best-fit  spectral model, a light curve is produced. The model light curve can be found in Figure \ref{fig:BolLC}; it provides a good fit to the bolometric light curve of SN\,2016jca, demonstrating that our density profile, ejecta mass, and \KE\ are consistent with both the light curve and spectral properties of the SN. It confirms that the \Nifs\ mass produced in the explosion was 0.27\,\Msun, and therefore \Nifs\ decay was luminous enough to power the light curve. 
The model light curve predicts that the observed bolometric magnitude of the SN would be 28\,mag at day +265, when the spectrum of the host galaxy was obtained. The ejecta of the SN at this time will have been too faint to be observed by the VLT; hence, no signs of it appear in the late-time observation, as discussed above.  

\section{Discussion}
The full opening angle of GRB\,161219B ($42^\circ \pm 3^\circ$) is consistent with only 5 out of 68 previously analyzed 
long-duration GRBs, whose full opening-angle distribution peaks at 
$10^\circ \pm 2^\circ$ \citep{Fong14}. 
It is also significantly larger than that of other low-redshift 
GRBs which were associated with SNe, whose opening angles can be accurately determined, except for GRBs\,980425 and 
031203, which showed no apparent jet break (see Table \ref{table:GRBSNprop}).  
However, correcting the isotropic 
energy of GRB\,161219B ($1.6 \times 10^{50}$\,erg; Frederiks et al. 2017) for the solid 
angle subtended by the jet aperture, an intrinsic jet energy of 
$\sim 10^{49}$\,erg is obtained, which is similar to all other events after
correcting for collimation.  
The large opening angle may be
the consequence of the widening of the jet after it emerges from the star. In
fact, it is unlikely that a jet can carve a ``cone'' of $\sim 42^\circ$ in the
stellar envelope, as this would require a very large amount of energy. Initially
jets are probably well collimated, and they can widen only after they break out of
the stellar surface \citep{Mizuta13}. So, in the case of GRB\,161219B, a larger
opening angle would correspond to a smaller amount of energy per solid angle, with
the total remaining roughly constant. In GRB-SNe with smaller jet opening angles
the widening may have been less significant, or we may have seen the events more
on axis if the energy was not uniformly distributed over angle after the jets
widened.\footnote{This scenario may also explain why GRB\,980425 had such a
small isotropic \KE\ ($\sim 10^{48}$\,erg): if the event was observed 10$^\circ$--15$^\circ$
off axis, as suggested by nebular spectroscopy of SN\,1998bw
\citep{Maeda02}, the jet may have spread out significantly. 
If the
energy was not uniformly distributed after jet widening, but more peaked toward
the jet axis, the value of \KE(iso) obtained from the energy observed several
degrees off axis could significantly underestimate the real \KE.)} It should
also be noted that the energy observed in the GRB cannot be the energy carried by the
material that punches a hole in the star: that energy is used up. Rather, the GRB is made
by new material that comes out unimpeded. The energy of the GRB depends on the
length of time over which the engine was still active after the jet broke
out \citep{Lazzati12}. 

We have plotted the location of SN\,2016jca on the correlations of ejected mass, \KE, mass of progenitor, and \Nifs\ mass; see Figure \ref{fig:corelations}. The location of SN\,2016jca is similar to that of other GRB-SNe: it has a \KE/$M_{\rm {ej}}$ ratio of \ab6, consistent with  other GRB-SNe. SN\,2016jca has an ejected mass of \ab6.5\,\Msun, lower than that of other GRB-SNe but larger than less-energetic events that did not harbour GRBs.
Revealing that GRB-SNe have ejecta masses which range from 6 to 11\,\Msun. Finally, the mass of the progenitor is consistent with an event creating a black hole rather than a neutron star, similar to all other GRB-SNe.

\citet{Cano17} present an analysis of SN\,2016jca; they, too, claim that the luminosity of the SN is powered by radioactive \Nifs\ decay.  Their best-fit parameters from fitting the SN light curve were that SN\,2016jca had \Mej = 5.8\,\Msun, \KE = $5 \times 10^{52}$\,erg, and synthesized 0.22 \Msun of \Nifs.   These values are similar to those obtained from our work, and any discrepancies are likely to be due to the simplistic assumptions they use.  As explained extensively by \citet{Mazzali17}, the method implemented by \citet{Cano17} depends on too many simplifying assumptions and therefore cannot yield reliable results. Incidentally, we note that \citet{Cano17} are confused about the role of a magnetar in our interpretation of SN\,2016jca: they make it appear as if we suggest that the magnetar is the source of luminosity. Magnetar energy may be released to produce explosion kinetic energy, and the accompanying nucleosynthesis may lead to the synthesis of $^{56}$Ni.  However, in our model, radioactivity is the only source of the luminosity. The light curve of SN\,2016jca and all other GRB-SNe can be explained simply by radioactive decay input, and this is consistent with nebular-phase observations that can be analyzed to yield the Fe mass (currently available only for SN\,1998bw and SN\,2006aj; \citealt{Mazzali01,Maeda07,Mazzali07}).

Most well-studied GRB-SNe have a remarkably narrow distribution of properties such as
luminosity ($\sim -18.7 \pm 0.2$\,mag), ejecta mass ($\sim 10\pm 2$\,\Msun), and 
\KE\  ($\sim 4 \pm 1 \times 10^{52}$\,erg\,s$^{-1}$ before correction for
asphericity); see \citet{Mazzali17}. 
However, SN\,2016jca has a low ejecta mass for a GRB-SN, demonstrating that there is diversity among GRB-SNe.
In fact, SN\,2016jca may have had a progenitor which came from a zero-age main sequence (ZAMS) star of $\sim 25$--35\,\Msun. 
Regardless, a normal neutrino-driven SN is unlikely to reach high \KE\ and to 
produce several 0.1\,\Msun\ of \Nifs.
Based on the similarity of the SN \KE\  [(1--2) $\times 10^{52}$\,erg after 
correcting for asphericity] to the rotational energy of a millisecond pulsar, it 
has been suggested that magnetars (highly magnetised and rapidly spinning neutron 
stars) could be viable central engines \citep{Mazzali06,Metzger2011,Mazzali14}.  
If rotational energy is released very rapidly, magnetars could provide \KE\ to 
the SN ejecta and contribute to nucleosynthesis, leading to the overproduction of 
\Nifs. Magnetars may also produce GRB jets \citep{Uzdensky07}, but it is unclear 
whether the material in the jet can reach the observed high Lorentz factors.\footnote{Models have been proposed where the magnetar powers the light
curve of superluminous SNe via interaction. This is NOT the scenario we envision here. If we
used the magnetar energy to power the light curve, a long spindown time would need
to be used, and there would not be enough energy at the earliest times to power 
the SN \KE, leaving unanswered the question of how the SN acquires its \KE.}

On the other hand, if the remnant is a black hole, stellar material may accrete 
onto the black hole, which launches a pair of jets along the rotational axis.  
In this ``collapsar'' scenario the jets may break out of the star and produce  
GRBs. It is unclear whether the jets can also deposit the energy required to 
explode the star violently: only (2--3) $\times 10^{51}$\,erg of energy are required 
for the jet to penetrate the stellar envelope, and thereafter no coupling occurs 
\citep{Lazzati13}. The large SN \KE\ may be produced in a broader outflow driven
by a disc wind, where \Nifs\ may also be synthesised. Some \Nifs\ may also be
created if the collapse to a black hole is preceded by a short-lived 
proto-neutron-star phase, outside which energy is deposited.  

One possibility is that the star first collapses to a magnetar creating a
neutrino-driven explosion aided by magnetar rotational energy in achieving large
\KE\ and \Nifs\ mass to power the SN ejecta and light curve. To have any
significant energy input at early times, the spin period ($P$) must be less 
than a few ms and the magnetic field ($B$) must be $\sim 1e15$\,G. 
The neutron star may quickly collapse to a black hole if it spins down
rapidly, and accretion onto the black hole may create the GRB jets and the
high-velocity \Nifs.

Regardless of the nature of the central engine, the fact that SN\,2016jca had significantly suppressed flux at shorter wavelengths compared to 
other GRB-SNe indicates that it had a 
significant amount of \Nifs\ at high velocity, which may imply that the explosion was
aspherical and that our viewing angle must have been close to the jet axis.  All
of this high-velocity 
\Nifs\ ($\sim 0.03$\,\Msun\ of \Nifs\ above 10,000\,\kms)  could not have been made directly in
the jet. Alternatively, 
high-velocity \Nifs\ could have been produced in the
region surrounding the jet as it pushed through the star, dredged up from the
centre of the explosion in an broad outflow next to the jet, or produced in a disc
wind around an accreting black hole in the collapsar model. 

Recent work has also claimed evidence for high-velocity \Nifs\ in GRB-SN  2017iuk and an early cocoon generated by the GRB jet \citep{Izzo19}. However, we note that spectral models of the radial dependence of \Nifs\ in a GRB-SN, comparison to observations, and a very similar modelling code were first presented and suggested in the original preprint of the present paper\footnote{https://arxiv.org/abs/1901.05500}. This preprint was first available in February 2017, two years before the work of \citet{Izzo19}.

\section{Conclusion}
We have presented a multiwavelength analysis of the afterglow of GRB\,161219B and the associated SN\,2016jca. 

A refreshed  shock with a luminosity $L(t) \propto t^{-0.85}$, that slows down the time decay of the afterglow, is virtually required by the observed time-decay indices of the X-ray light curve, and it helps solve the mild inconsistency in the fireball closure relationship  noted by  \citet{Cano17}.   Our estimate of the steepening time of the X-ray light curve yields a break time of $\sim26$\,d that is formally consistent with that obtained by  \citet{Cano17}   ($t_b \approx 30$\,d), but would cause the SN luminosity to be too low to be compatible with the observed spectra.  Therefore, we assumed a break time of $t_b =  13 \pm 2$\,d, which corresponds to the lower end of the confidence interval  we determined for this parameter. The corresponding full-jet opening angle is $42^\circ \pm 3^\circ$.
This turns out to be the widest opening angle that has been calculated for a low-redshift GRB-SNe. 
However, after correcting for collimation the GRB energy was $10^{49}$\,erg, which is comparable to all previous low-redshift events. This suggests that their underlying engine must be somewhat invariant.  

The bolometric light curve of SN\,2016jca peaked $12 \pm 2$\,d after explosion at $-18.2 \pm 0.1$\,mag.
The SN  synthesised roughly $0.27 \pm 0.05$\,\Msun\ of \Nifs, but had a more quickly evolving light curve with respect to other GRB-SNe. 
After correction for afterglow contamination, the spectra of SN\,2016jca show
suppressed flux at shorter wavelengths with respect to other GRB-SNe. This indicates that there is increased line blanketing in the ejecta of SN\,2016jca with respect to other events. 

To examine the physics of the SN further we turn to radiative-transfer models. 
Our models determined that the SN had an ejecta mass of $6.5 \pm 1.5$\,\Msun\ and 
\KE\ $=4\pm0.8\times10^{52}$\,erg, or (1--3) $\times10^{52}$\,erg
when corrected for asphericity. The models demonstrated 
that at 3.73\,d the photospheric velocity
of SN\,2016jca was $35,000 \pm 7200$\,\kms, making this the largest photospheric velocity
of any stripped-envelope SN. Furthermore, in our models we need a large (0.4\% or 0.03\,\Msun) \Nifs\ abundance 
at high velocities to provide opacity and line blanketing in the blue. In fact, the models
favour a decreasing \Nifs\ abundances as a function of decreasing velocity. This suggests that SN\,2016jca was a highly aspherical explosion viewed close to on-axis. 

The similarities between the physical properties of GRB-SNe, such as the energy of the GRB when corrected for collimation (\ab$10^{49-50}$\,erg),
the \KE\ of the SN when corrected for asphericity [(1--2)$ \times10^{52}$\,erg], the average rise time (\ab13\,d), and the \Nifs\ mass (\ab$0.27\pm0.05$), demonstrate that these events must have very similar progenitor 
systems. They could possibly be Wolf-Rayet stars with ZAMS masses of 25--50\,\Msun. 
Furthermore, their central engines must be similar, and there are arguments for and against both the collapsar and magnetar scenarios. 
Regardless of the nature of central engine, our early-time 
spectra allowed us to place tight constraints on the \KE\ of SN\,2016jca. The
extremely small ratio ($<0.1$\%) between the energy promptly emitted by
GRB\,161219B in gamma-rays and the kinetic energy of SN\,2016jca highlights the fact
that the bulk of the energy is carried by the SN and not the jet.

\section*{Acknowledgements} 

Chris Ashall acknowledges support provided by the National Science Foundation (NSF) under grant AST--1613472.
Work in this paper was based on observations made with ESO 
telescopes at the Paranal Observatory under programmes ID  098.D-0055(A), 098.D-0218(A), 298.D-5022(A), and 0100.D-0504(A).
M.S. and S.H. are supported by a generous grant (13261) from VILLUM FONDEN. M.S. is also supported by a project grant from the IRFD (Independent Research Fund Denmark).
A portion of this work was supported by a grant fron IDA (Danish Instrument Center for Danish Astrophysics).
The work of A.V.F. and W.Z. has been supported by the 
Christopher R. Redlich Fund, the TABASGO Foundation, and the Miller
Institute for Basic Research in Science (U.C. Berkeley).
S.R.O. acknowledges the support of the Leverhulme Trust. 
E.P. acknowledges Scuola Normale Superiore and INAF for support.  
A.F.V. thanks the Russian Science Foundation for grant 14-50-00043.
A.M. acknowledges support from the ASI INAF grant I/004/11/1.
A.J.C.T. acknowledges support from the Spanish Ministry Project AYA2015-71718-R. 
SW at UCSC was supported by the NASA Theory Program (NNX14AH34G)

Both the Liverpool Telescope and the GTC are operated on the  island of La Palma by Liverpool John Moores University in the Spanish Observatorio del Roque de los Muchachos of the Instituto de Astrofisica de Canarias with financial support from the UK Science and Technology Facilities Council.
Based on observations made with the TNG, operated on the island of La Palma by the Fundaci\'{o}n Galileo Galilei 
of the Instituto Nazionale di Astrofisica (INAF) at the Spanish Observatorio del Roque
de los Muchachos of the Instituto de Astrofisica de Canarias under programme $A32TAC_5$.
Some of the data presented herein were obtained at the W. M. Keck
Observatory, which is operated as a scientific partnership among the
California Institute of Technology, the University of California, and
the National Aeronautics and Space Administration (NASA); 
the observatory was made possible by the generous financial
support of the W. M. Keck Foundation.
We thank the Paranal Director for allocating Discretionary Time to this programme, the Paranal Science Operations team for their assistance, and T. Kr\"uhler for assistance with X-Shooter data reduction. 
This work made use of data supplied by the UK Swift Science Data Centre 
at the University of Leicester.
IRAF is the Image Reduction and Analysis Facility made available
to  the  astronomical  community  by  the  National  Optical  Astronomy
Observatories, which are operated by AURA, Inc., under contract with
the US National Science Foundation. It is available at
http://iraf.noao.edu. 

The Pan-STARRS1 Surveys (PS1) and the PS1 public science archive have been made possible through contributions by the Institute for Astronomy, the University of Hawaii, the Pan-STARRS Project Office, the Max-Planck Society and its participating institutes, the Max Planck Institute for Astronomy, Heidelberg and the Max Planck Institute for Extraterrestrial Physics, Garching, The Johns Hopkins University, Durham University, the University of Edinburgh, the Queen's University Belfast, the Harvard-Smithsonian Center for Astrophysics, the Las Cumbres Observatory Global Telescope Network Incorporated, the National Central University of Taiwan, the Space Telescope Science Institute, NASA grant NNX08AR22G issued through the Planetary Science Division of the NASA Science Mission Directorate, NSF grant AST-1238877, the University of Maryland, Eotvos Lorand University (ELTE), the Los Alamos National Laboratory, and the Gordon and Betty Moore Foundation.






\bibliographystyle{mnras}
\bibliography{BB}

\begin{thebibliography}{}
\makeatletter
\relax
\def\mn@urlcharsother{\let\do\@makeother \do\$\do\&\do\#\do\^\do\_\do\%\do\~}
\def\mn@doi{\begingroup\mn@urlcharsother \@ifnextchar [ {\mn@doi@}
  {\mn@doi@[]}}
\def\mn@doi@[#1]#2{\def\@tempa{#1}\ifx\@tempa\@empty \href
  {http://dx.doi.org/#2} {doi:#2}\else \href {http://dx.doi.org/#2} {#1}\fi
  \endgroup}
\def\mn@eprint#1#2{\mn@eprint@#1:#2::\@nil}
\def\mn@eprint@arXiv#1{\href {http://arxiv.org/abs/#1} {{\tt arXiv:#1}}}
\def\mn@eprint@dblp#1{\href {http://dblp.uni-trier.de/rec/bibtex/#1.xml}
  {dblp:#1}}
\def\mn@eprint@#1:#2:#3:#4\@nil{\def\@tempa {#1}\def\@tempb {#2}\def\@tempc
  {#3}\ifx \@tempc \@empty \let \@tempc \@tempb \let \@tempb \@tempa \fi \ifx
  \@tempb \@empty \def\@tempb {arXiv}\fi \@ifundefined
  {mn@eprint@\@tempb}{\@tempb:\@tempc}{\expandafter \expandafter \csname
  mn@eprint@\@tempb\endcsname \expandafter{\@tempc}}}

\bibitem[\protect\citeauthoryear{{Ashall} \& {Mazzali}}{{Ashall} \&
  {Mazzali}}{2019}]{Ashall19}
{Ashall} C.,  {Mazzali} P.~A.,  2019, preprint, \href
  {http://adsabs.harvard.edu/abs/2017arXiv170408298P} {} (\mn@eprint {arXiv}
  {submitted})

\bibitem[\protect\citeauthoryear{{Ashall}, {Mazzali}, {Bersier}, {Hachinger},
  {Phillips}, {Percival}, {James}  \& {Maguire}}{{Ashall}
  et~al.}{2014}]{Ashall14}
{Ashall} C.,  {Mazzali} P.,  {Bersier} D.,  {Hachinger} S.,  {Phillips} M.,
  {Percival} S.,  {James} P.,   {Maguire} K.,  2014, \mn@doi [\mnras]
  {10.1093/mnras/stu1995}, \href
  {http://adsabs.harvard.edu/abs/2014MNRAS.445.4427A} {445, 4427}

\bibitem[\protect\citeauthoryear{{Ashall}, {Mazzali}, {Pian}  \&
  {James}}{{Ashall} et~al.}{2016}]{Ashall16}
{Ashall} C.,  {Mazzali} P.~A.,  {Pian} E.,   {James} P.~A.,  2016, \mn@doi
  [\mnras] {10.1093/mnras/stw2114}, \href
  {http://adsabs.harvard.edu/abs/2016MNRAS.463.1891A} {463, 1891}

\bibitem[\protect\citeauthoryear{{Ashall} et~al.,}{{Ashall}
  et~al.}{2018}]{Ashall18}
{Ashall} C.,  et~al., 2018, \mn@doi [\mnras] {10.1093/mnras/sty632}, \href
  {http://adsabs.harvard.edu/abs/2018MNRAS.477..153A} {477, 153}

\bibitem[\protect\citeauthoryear{{Avni}}{{Avni}}{1976}]{Avni76}
{Avni} Y.,  1976, \mn@doi [\apj] {10.1086/154870}, \href
  {http://adsabs.harvard.edu/abs/1976ApJ...210..642A} {210, 642}

\bibitem[\protect\citeauthoryear{{Barnes}, {Duffell}, {Liu}, {Modjaz},
  {Bianco}, {Kasen}  \& {MacFadyen}}{{Barnes} et~al.}{2018}]{Barnes18}
{Barnes} J.,  {Duffell} P.~C.,  {Liu} Y.,  {Modjaz} M.,  {Bianco} F.~B.,
  {Kasen} D.,   {MacFadyen} A.~I.,  2018, \mn@doi [\apj]
  {10.3847/1538-4357/aabf84}, \href
  {http://adsabs.harvard.edu/abs/2018ApJ...860...38B} {860, 38}

\bibitem[\protect\citeauthoryear{{Breeveld}, {Landsman}, {Holland}, {Roming},
  {Kuin}  \& {Page}}{{Breeveld} et~al.}{2011}]{Breeveld11}
{Breeveld} A.~A.,  {Landsman} W.,  {Holland} S.~T.,  {Roming} P.,  {Kuin}
  N.~P.~M.,   {Page} M.~J.,  2011, in {McEnery} J.~E.,  {Racusin} J.~L.,
  {Gehrels} N.,  eds,  American Institute of Physics Conference Series Vol.
  1358, American Institute of Physics Conference Series. pp 373--376
  (\mn@eprint {arXiv} {1102.4717}), \mn@doi{10.1063/1.3621807}

\bibitem[\protect\citeauthoryear{{Bufano} et~al.,}{{Bufano}
  et~al.}{2012}]{Bufano2012}
{Bufano} F.,  et~al., 2012, \mn@doi [\apj] {10.1088/0004-637X/753/1/67}, \href
  {http://adsabs.harvard.edu/abs/2012ApJ...753...67B} {753, 67}

\bibitem[\protect\citeauthoryear{{Cano} et~al.,}{{Cano} et~al.}{2017}]{Cano17}
{Cano} Z.,  et~al., 2017, \mn@doi [\aap] {10.1051/0004-6361/201731005}, \href
  {http://adsabs.harvard.edu/abs/2017A%26A...605A.107C} {605, A107}

\bibitem[\protect\citeauthoryear{{Cappellaro}, {Mazzali}, {Benetti},
  {Danziger}, {Turatto}, {della Valle}  \& {Patat}}{{Cappellaro}
  et~al.}{1997}]{Cappellaro97}
{Cappellaro} E.,  {Mazzali} P.~A.,  {Benetti} S.,  {Danziger} I.~J.,  {Turatto}
  M.,  {della Valle} M.,   {Patat} F.,  1997, \aap, \href
  {http://adsabs.harvard.edu/abs/1997A%26A...328..203C} {328, 203}

\bibitem[\protect\citeauthoryear{{Cardelli}, {Clayton}  \& {Mathis}}{{Cardelli}
  et~al.}{1989}]{Cardelli89}
{Cardelli} J.~A.,  {Clayton} G.~C.,   {Mathis} J.~S.,  1989, \mn@doi [\apj]
  {10.1086/167900}, \href {http://adsabs.harvard.edu/abs/1989ApJ...345..245C}
  {345, 245}

\bibitem[\protect\citeauthoryear{{D'Ai} et~al.,}{{D'Ai}
  et~al.}{2016}]{GCN20296}
{D'Ai} A.,  et~al., 2016, GRB Coordinates Network, \href
  {http://adsabs.harvard.edu/abs/2016GCN..20296...1D} {20296}

\bibitem[\protect\citeauthoryear{{D'Elia} et~al.,}{{D'Elia}
  et~al.}{2015}]{Delia15}
{D'Elia} V.,  et~al., 2015, \mn@doi [\aap] {10.1051/0004-6361/201425381}, \href
  {http://adsabs.harvard.edu/abs/2015A%26A...577A.116D} {577, A116}

\bibitem[\protect\citeauthoryear{{D'Odorico} et~al.,}{{D'Odorico}
  et~al.}{2006}]{DOdorico06}
{D'Odorico} S.,  et~al., 2006, in Society of Photo-Optical Instrumentation
  Engineers (SPIE) Conference Series. p. 626933, \mn@doi{10.1117/12.672969}

\bibitem[\protect\citeauthoryear{{Dai} \& {Cheng}}{{Dai} \&
  {Cheng}}{2001}]{Dai01}
{Dai} Z.~G.,  {Cheng} K.~S.,  2001, \mn@doi [\apjl] {10.1086/323566}, \href
  {http://adsabs.harvard.edu/abs/2001ApJ...558L.109D} {558, L109}

\bibitem[\protect\citeauthoryear{{Deng}, {Tominaga}, {Mazzali}, {Maeda}  \&
  {Nomoto}}{{Deng} et~al.}{2005}]{Deng05}
{Deng} J.,  {Tominaga} N.,  {Mazzali} P.~A.,  {Maeda} K.,   {Nomoto} K.,  2005,
  \mn@doi [\apj] {10.1086/429284}, \href
  {http://adsabs.harvard.edu/abs/2005ApJ...624..898D} {624, 898}

\bibitem[\protect\citeauthoryear{{Dexter} \& {Kasen}}{{Dexter} \&
  {Kasen}}{2013}]{Dexter13}
{Dexter} J.,  {Kasen} D.,  2013, \mn@doi [\apj] {10.1088/0004-637X/772/1/30},
  \href {http://adsabs.harvard.edu/abs/2013ApJ...772...30D} {772, 30}

\bibitem[\protect\citeauthoryear{{Drout} et~al.,}{{Drout}
  et~al.}{2011}]{Drout11}
{Drout} M.~R.,  et~al., 2011, \mn@doi [\apj] {10.1088/0004-637X/741/2/97},
  \href {http://adsabs.harvard.edu/abs/2011ApJ...741...97D} {741, 97}

\bibitem[\protect\citeauthoryear{{Ferrero} et~al.,}{{Ferrero}
  et~al.}{2006}]{Ferrero06}
{Ferrero} P.,  et~al., 2006, \mn@doi [\aap] {10.1051/0004-6361:20065530}, \href
  {http://adsabs.harvard.edu/abs/2006A%26A...457..857F} {457, 857}

\bibitem[\protect\citeauthoryear{{Filippenko}}{{Filippenko}}{1982}]{Filippenko82}
{Filippenko} A.~V.,  1982, \mn@doi [\pasp] {10.1086/131052}, \href
  {http://adsabs.harvard.edu/abs/1982PASP...94..715F} {94, 715}

\bibitem[\protect\citeauthoryear{{Filippenko}}{{Filippenko}}{1997}]{Filippenko97}
{Filippenko} A.~V.,  1997, \mn@doi [\araa] {10.1146/annurev.astro.35.1.309},
  \href {http://adsabs.harvard.edu/abs/1997ARA%26A..35..309F} {35, 309}

\bibitem[\protect\citeauthoryear{{Fong} et~al.,}{{Fong} et~al.}{2014}]{Fong14}
{Fong} W.,  et~al., 2014, \mn@doi [\apj] {10.1088/0004-637X/780/2/118}, \href
  {http://adsabs.harvard.edu/abs/2014ApJ...780..118F} {780, 118}

\bibitem[\protect\citeauthoryear{{Frederiks} et~al.,}{{Frederiks}
  et~al.}{2016}]{Frederiks17}
{Frederiks} D.,  et~al., 2016, GRB Coordinates Network, \href
  {http://adsabs.harvard.edu/abs/2016GCN..20323...1F} {20323}

\bibitem[\protect\citeauthoryear{{Fukugita}, {Shimasaku}  \&
  {Ichikawa}}{{Fukugita} et~al.}{1995}]{Fukugita95}
{Fukugita} M.,  {Shimasaku} K.,   {Ichikawa} T.,  1995, \mn@doi [\pasp]
  {10.1086/133643}, \href {http://adsabs.harvard.edu/abs/1995PASP..107..945F}
  {107, 945}

\bibitem[\protect\citeauthoryear{{Galama} et~al.,}{{Galama}
  et~al.}{1998}]{Galama98}
{Galama} T.~J.,  et~al., 1998, \mn@doi [\nat] {10.1038/27150}, \href
  {http://adsabs.harvard.edu/abs/1998Natur.395..670G} {395, 670}

\bibitem[\protect\citeauthoryear{{Hjorth} et~al.,}{{Hjorth}
  et~al.}{2003}]{Hjorth03}
{Hjorth} J.,  et~al., 2003, \mn@doi [\nat] {10.1038/nature01750}, \href
  {http://adsabs.harvard.edu/abs/2003Natur.423..847H} {423, 847}

\bibitem[\protect\citeauthoryear{{Horne}}{{Horne}}{1986}]{Horne86}
{Horne} K.,  1986, \mn@doi [\pasp] {10.1086/131801}, \href
  {http://adsabs.harvard.edu/abs/1986PASP...98..609H} {98, 609}

\bibitem[\protect\citeauthoryear{{Iwamoto} et~al.,}{{Iwamoto}
  et~al.}{1998}]{Iwamoto98}
{Iwamoto} K.,  et~al., 1998, \mn@doi [\nat] {10.1038/27155}, \href
  {http://adsabs.harvard.edu/abs/1998Natur.395..672I} {395, 672}

\bibitem[\protect\citeauthoryear{{Izzo} et~al.,}{{Izzo} et~al.}{2019}]{Izzo19}
{Izzo} L.,  et~al., 2019, \mn@doi [\nat] {10.1038/s41586-018-0826-3}, \href
  {https://ui.adsabs.harvard.edu/abs/2019Natur.565..324I} {565, 324}

\bibitem[\protect\citeauthoryear{{Janka}}{{Janka}}{2012}]{Janka12}
{Janka} H.-T.,  2012, \mn@doi [Annual Review of Nuclear and Particle Science]
  {10.1146/annurev-nucl-102711-094901}, \href
  {http://adsabs.harvard.edu/abs/2012ARNPS..62..407J} {62, 407}

\bibitem[\protect\citeauthoryear{{Kasen} \& {Bildsten}}{{Kasen} \&
  {Bildsten}}{2010}]{Kasen10}
{Kasen} D.,  {Bildsten} L.,  2010, \mn@doi [\apj]
  {10.1088/0004-637X/717/1/245}, \href
  {http://adsabs.harvard.edu/abs/2010ApJ...717..245K} {717, 245}

\bibitem[\protect\citeauthoryear{{Kinney}, {Calzetti}, {Bohlin}, {McQuade},
  {Storchi-Bergmann}  \& {Schmitt}}{{Kinney} et~al.}{1996}]{Kinney96}
{Kinney} A.~L.,  {Calzetti} D.,  {Bohlin} R.~C.,  {McQuade} K.,
  {Storchi-Bergmann} T.,   {Schmitt} H.~R.,  1996, \mn@doi [\apj]
  {10.1086/177583}, \href {http://adsabs.harvard.edu/abs/1996ApJ...467...38K}
  {467, 38}

\bibitem[\protect\citeauthoryear{{Lazzati}, {Morsony}, {Blackwell}  \&
  {Begelman}}{{Lazzati} et~al.}{2012}]{Lazzati12}
{Lazzati} D.,  {Morsony} B.~J.,  {Blackwell} C.~H.,   {Begelman} M.~C.,  2012,
  \mn@doi [\apj] {10.1088/0004-637X/750/1/68}, \href
  {http://adsabs.harvard.edu/abs/2012ApJ...750...68L} {750, 68}

\bibitem[\protect\citeauthoryear{{Lazzati}, {Villeneuve},
  {L{\'o}pez-C{\'a}mara}, {Morsony}  \& {Perna}}{{Lazzati}
  et~al.}{2013}]{Lazzati13}
{Lazzati} D.,  {Villeneuve} M.,  {L{\'o}pez-C{\'a}mara} D.,  {Morsony} B.~J.,
  {Perna} R.,  2013, \mn@doi [\mnras] {10.1093/mnras/stt1705}, \href
  {http://adsabs.harvard.edu/abs/2013MNRAS.436.1867L} {436, 1867}

\bibitem[\protect\citeauthoryear{{Lucy}}{{Lucy}}{1999}]{Lucy99}
{Lucy} L.~B.,  1999, \aap, \href
  {http://adsabs.harvard.edu/abs/1999A%26A...345..211L} {345, 211}

\bibitem[\protect\citeauthoryear{{Lyman}, {Bersier}, {James}, {Mazzali},
  {Eldridge}, {Fraser}  \& {Pian}}{{Lyman} et~al.}{2016}]{Lyman16}
{Lyman} J.~D.,  {Bersier} D.,  {James} P.~A.,  {Mazzali} P.~A.,  {Eldridge}
  J.~J.,  {Fraser} M.,   {Pian} E.,  2016, \mn@doi [\mnras]
  {10.1093/mnras/stv2983}, \href
  {http://adsabs.harvard.edu/abs/2016MNRAS.457..328L} {457, 328}

\bibitem[\protect\citeauthoryear{{MacFadyen} \& {Woosley}}{{MacFadyen} \&
  {Woosley}}{1999}]{MacFadyen99}
{MacFadyen} A.~I.,  {Woosley} S.~E.,  1999, \mn@doi [\apj] {10.1086/307790},
  \href {http://adsabs.harvard.edu/abs/1999ApJ...524..262M} {524, 262}

\bibitem[\protect\citeauthoryear{{Maeda}, {Nakamura}, {Nomoto}, {Mazzali},
  {Patat}  \& {Hachisu}}{{Maeda} et~al.}{2002}]{Maeda02}
{Maeda} K.,  {Nakamura} T.,  {Nomoto} K.,  {Mazzali} P.~A.,  {Patat} F.,
  {Hachisu} I.,  2002, \mn@doi [\apj] {10.1086/324487}, \href
  {http://adsabs.harvard.edu/abs/2002ApJ...565..405M} {565, 405}

\bibitem[\protect\citeauthoryear{{Maeda} et~al.,}{{Maeda}
  et~al.}{2007}]{Maeda07}
{Maeda} K.,  et~al., 2007, \mn@doi [\apj] {10.1086/513564}, \href
  {https://ui.adsabs.harvard.edu/abs/2007ApJ...658L...5M} {658, L5}

\bibitem[\protect\citeauthoryear{{Malesani} et~al.,}{{Malesani}
  et~al.}{2004}]{Malesani04}
{Malesani} D.,  et~al., 2004, \mn@doi [\apjl] {10.1086/422684}, \href
  {http://adsabs.harvard.edu/abs/2004ApJ...609L...5M} {609, L5}

\bibitem[\protect\citeauthoryear{{Matheson} et~al.,}{{Matheson}
  et~al.}{2003}]{Matheson03}
{Matheson} T.,  et~al., 2003, \mn@doi [\apj] {10.1086/379228}, \href
  {http://adsabs.harvard.edu/abs/2003ApJ...599..394M} {599, 394}

\bibitem[\protect\citeauthoryear{{Mazzali} \& {Lucy}}{{Mazzali} \&
  {Lucy}}{1993}]{Mazzali93}
{Mazzali} P.~A.,  {Lucy} L.~B.,  1993, \aap, \href
  {http://adsabs.harvard.edu/abs/1993A%26A...279..447M} {279, 447}

\bibitem[\protect\citeauthoryear{{Mazzali}, {Iwamoto}  \& {Nomoto}}{{Mazzali}
  et~al.}{2000}]{Mazzali00}
{Mazzali} P.~A.,  {Iwamoto} K.,   {Nomoto} K.,  2000, \mn@doi [\apj]
  {10.1086/317808}, \href {http://adsabs.harvard.edu/abs/2000ApJ...545..407M}
  {545, 407}

\bibitem[\protect\citeauthoryear{{Mazzali}, {Nomoto}, {Patat}  \&
  {Maeda}}{{Mazzali} et~al.}{2001}]{Mazzali01}
{Mazzali} P.~A.,  {Nomoto} K.,  {Patat} F.,   {Maeda} K.,  2001, \mn@doi [\apj]
  {10.1086/322420}, \href
  {https://ui.adsabs.harvard.edu/abs/2001ApJ...559.1047M} {559, 1047}

\bibitem[\protect\citeauthoryear{{Mazzali} et~al.,}{{Mazzali}
  et~al.}{2002}]{Mazzali02}
{Mazzali} P.~A.,  et~al., 2002, \mn@doi [\apjl] {10.1086/341504}, \href
  {http://adsabs.harvard.edu/abs/2002ApJ...572L..61M} {572, L61}

\bibitem[\protect\citeauthoryear{{Mazzali} et~al.,}{{Mazzali}
  et~al.}{2006}]{Mazzali06}
{Mazzali} P.~A.,  et~al., 2006, \mn@doi [\nat] {10.1038/nature05081}, \href
  {http://adsabs.harvard.edu/abs/2006Natur.442.1018M} {442, 1018}

\bibitem[\protect\citeauthoryear{{Mazzali} et~al.,}{{Mazzali}
  et~al.}{2007}]{Mazzali07}
{Mazzali} P.~A.,  et~al., 2007, \mn@doi [\apj] {10.1086/517912}, \href
  {https://ui.adsabs.harvard.edu/abs/2007ApJ...661..892M} {661, 892}

\bibitem[\protect\citeauthoryear{{Mazzali} et~al.,}{{Mazzali}
  et~al.}{2008}]{Mazzali08a}
{Mazzali} P.~A.,  et~al., 2008, \mn@doi [Science] {10.1126/science.1158088},
  \href {http://adsabs.harvard.edu/abs/2008Sci...321.1185M} {321, 1185}

\bibitem[\protect\citeauthoryear{{Mazzali}, {Walker}, {Pian}, {Tanaka},
  {Corsi}, {Hattori}  \& {Gal-Yam}}{{Mazzali} et~al.}{2013}]{Mazzali13}
{Mazzali} P.~A.,  {Walker} E.~S.,  {Pian} E.,  {Tanaka} M.,  {Corsi} A.,
  {Hattori} T.,   {Gal-Yam} A.,  2013, \mn@doi [\mnras] {10.1093/mnras/stt605},
  \href {http://adsabs.harvard.edu/abs/2013MNRAS.432.2463M} {432, 2463}

\bibitem[\protect\citeauthoryear{{Mazzali}, {McFadyen}, {Woosley}, {Pian}  \&
  {Tanaka}}{{Mazzali} et~al.}{2014}]{Mazzali14}
{Mazzali} P.~A.,  {McFadyen} A.~I.,  {Woosley} S.~E.,  {Pian} E.,   {Tanaka}
  M.,  2014, \mn@doi [\mnras] {10.1093/mnras/stu1124}, \href
  {http://adsabs.harvard.edu/abs/2014MNRAS.443...67M} {443, 67}

\bibitem[\protect\citeauthoryear{{Mazzali}, {Sauer}, {Pian}, {Deng},
  {Prentice}, {Ben Ami}, {Taubenberger}  \& {Nomoto}}{{Mazzali}
  et~al.}{2017}]{Mazzali17}
{Mazzali} P.~A.,  {Sauer} D.~N.,  {Pian} E.,  {Deng} J.,  {Prentice} S.,  {Ben
  Ami} S.,  {Taubenberger} S.,   {Nomoto} K.,  2017, \mn@doi [\mnras]
  {10.1093/mnras/stx992}, \href
  {http://adsabs.harvard.edu/abs/2017MNRAS.469.2498M} {469, 2498}

\bibitem[\protect\citeauthoryear{{Metzger}, {Giannios}, {Thompson},
  {Bucciantini}  \& {Quataert}}{{Metzger} et~al.}{2011}]{Metzger2011}
{Metzger} B.~D.,  {Giannios} D.,  {Thompson} T.~A.,  {Bucciantini} N.,
  {Quataert} E.,  2011, \mn@doi [\mnras] {10.1111/j.1365-2966.2011.18280.x},
  \href {http://adsabs.harvard.edu/abs/2011MNRAS.413.2031M} {413, 2031}

\bibitem[\protect\citeauthoryear{{Mingo} et~al.,}{{Mingo}
  et~al.}{2016}]{Mingo16}
{Mingo} B.,  et~al., 2016, GRB Coordinates Network, \href
  {https://ui.adsabs.harvard.edu/abs/2016GCN.20298....1M} {20298, 1}

\bibitem[\protect\citeauthoryear{{Mizuta} \& {Ioka}}{{Mizuta} \&
  {Ioka}}{2013}]{Mizuta13}
{Mizuta} A.,  {Ioka} K.,  2013, \mn@doi [\apj] {10.1088/0004-637X/777/2/162},
  \href {http://adsabs.harvard.edu/abs/2013ApJ...777..162M} {777, 162}

\bibitem[\protect\citeauthoryear{{Modigliani} et~al.,}{{Modigliani}
  et~al.}{2010}]{Modigliani10}
{Modigliani} A.,  et~al., 2010, in Observatory Operations: Strategies,
  Processes, and Systems III. p. 773728, \mn@doi{10.1117/12.857211}

\bibitem[\protect\citeauthoryear{{Oates} et~al.,}{{Oates}
  et~al.}{2009}]{Oates09}
{Oates} S.~R.,  et~al., 2009, \mn@doi [\mnras]
  {10.1111/j.1365-2966.2009.14544.x}, \href
  {http://adsabs.harvard.edu/abs/2009MNRAS.395..490O} {395, 490}

\bibitem[\protect\citeauthoryear{{Oke} et~al.,}{{Oke} et~al.}{1995}]{Oke95}
{Oke} J.~B.,  et~al., 1995, \mn@doi [\pasp] {10.1086/133562}, \href
  {http://adsabs.harvard.edu/abs/1995PASP..107..375O} {107, 375}

\bibitem[\protect\citeauthoryear{{Pian} et~al.,}{{Pian} et~al.}{2006}]{Pian06}
{Pian} E.,  et~al., 2006, \mn@doi [\nat] {10.1038/nature05082}, \href
  {http://adsabs.harvard.edu/abs/2006Natur.442.1011P} {442, 1011}

\bibitem[\protect\citeauthoryear{{Pian}, {Palazzi}  \& {Perley}}{{Pian}
  et~al.}{2016}]{Pian17TNS}
{Pian} E.,  {Palazzi} E.,   {Perley} D.,  2016, Transient Name Server Discovery
  Report, \href {http://adsabs.harvard.edu/abs/2016TNSTR1090....1P} {1090}

\bibitem[\protect\citeauthoryear{{Piran}, {Nakar}, {Mazzali}  \&
  {Pian}}{{Piran} et~al.}{2017}]{Piran2017}
{Piran} T.,  {Nakar} E.,  {Mazzali} P.,   {Pian} E.,  2017, preprint, \href
  {http://adsabs.harvard.edu/abs/2017arXiv170408298P} {} (\mn@eprint {arXiv}
  {1704.08298})

\bibitem[\protect\citeauthoryear{{Poole} et~al.,}{{Poole}
  et~al.}{2008}]{Poole08}
{Poole} T.~S.,  et~al., 2008, \mn@doi [\mnras]
  {10.1111/j.1365-2966.2007.12563.x}, \href
  {http://adsabs.harvard.edu/abs/2008MNRAS.383..627P} {383, 627}

\bibitem[\protect\citeauthoryear{{Prentice} \& {Mazzali}}{{Prentice} \&
  {Mazzali}}{2017}]{Prentice17}
{Prentice} S.~J.,  {Mazzali} P.~A.,  2017, \mn@doi [\mnras]
  {10.1093/mnras/stx980}, \href
  {http://adsabs.harvard.edu/abs/2017MNRAS.469.2672P} {469, 2672}

\bibitem[\protect\citeauthoryear{{Prentice} et~al.,}{{Prentice}
  et~al.}{2016}]{Prentice16}
{Prentice} S.~J.,  et~al., 2016, \mn@doi [\mnras] {10.1093/mnras/stw299}, \href
  {http://adsabs.harvard.edu/abs/2016MNRAS.458.2973P} {458, 2973}

\bibitem[\protect\citeauthoryear{{Prentice} et~al.,}{{Prentice}
  et~al.}{2018}]{Prentice18}
{Prentice} S.~J.,  et~al., 2018, \mn@doi [\mnras] {10.1093/mnras/sty1223},
  \href {http://adsabs.harvard.edu/abs/2018MNRAS.478.4162P} {478, 4162}

\bibitem[\protect\citeauthoryear{{Prentice} et~al.,}{{Prentice}
  et~al.}{2019}]{Prentice19}
{Prentice} S.~J.,  et~al., 2019, \mn@doi [\mnras] {10.1093/mnras/sty3399},
  \href {https://ui.adsabs.harvard.edu/abs/2019MNRAS.485.1559P} {485, 1559}

\bibitem[\protect\citeauthoryear{{Rhoads}}{{Rhoads}}{1999}]{Rhoads99}
{Rhoads} J.~E.,  1999, \mn@doi [\apj] {10.1086/307907}, \href
  {http://adsabs.harvard.edu/abs/1999ApJ...525..737R} {525, 737}

\bibitem[\protect\citeauthoryear{{Richmond} et~al.,}{{Richmond}
  et~al.}{1996}]{Richmond96}
{Richmond} M.~W.,  et~al., 1996, \mn@doi [\aj] {10.1086/117785}, \href
  {http://adsabs.harvard.edu/abs/1996AJ....111..327R} {111, 327}

\bibitem[\protect\citeauthoryear{{Sari}, {Piran}  \& {Narayan}}{{Sari}
  et~al.}{1998}]{Sari98}
{Sari} R.,  {Piran} T.,   {Narayan} R.,  1998, \mn@doi [\apjl]
  {10.1086/311269}, \href {http://adsabs.harvard.edu/abs/1998ApJ...497L..17S}
  {497, L17}

\bibitem[\protect\citeauthoryear{{Sari}, {Piran}  \& {Halpern}}{{Sari}
  et~al.}{1999}]{Sari99}
{Sari} R.,  {Piran} T.,   {Halpern} J.~P.,  1999, \mn@doi [\apjl]
  {10.1086/312109}, \href {http://adsabs.harvard.edu/abs/1999ApJ...519L..17S}
  {519, L17}

\bibitem[\protect\citeauthoryear{{Schlafly} \& {Finkbeiner}}{{Schlafly} \&
  {Finkbeiner}}{2011}]{Schlafly11}
{Schlafly} E.~F.,  {Finkbeiner} D.~P.,  2011, \mn@doi [\apj]
  {10.1088/0004-637X/737/2/103}, \href
  {http://adsabs.harvard.edu/abs/2011ApJ...737..103S} {737, 103}

\bibitem[\protect\citeauthoryear{{Stanek} et~al.,}{{Stanek}
  et~al.}{2003}]{Stanek03}
{Stanek} K.~Z.,  et~al., 2003, \mn@doi [\apjl] {10.1086/376976}, \href
  {http://adsabs.harvard.edu/abs/2003ApJ...591L..17S} {591, L17}

\bibitem[\protect\citeauthoryear{{Steele} et~al.,}{{Steele}
  et~al.}{2004}]{Steele04}
{Steele} I.~A.,  et~al., 2004, in {Oschmann} Jr. J.~M.,  ed.,  \procspie Vol.
  5489, Ground-based Telescopes. pp 679--692, \mn@doi{10.1117/12.551456}

\bibitem[\protect\citeauthoryear{{Tanaka} et~al.,}{{Tanaka}
  et~al.}{2009}]{Tanaka09}
{Tanaka} M.,  et~al., 2009, \mn@doi [\apj] {10.1088/0004-637X/700/2/1680},
  \href {http://adsabs.harvard.edu/abs/2009ApJ...700.1680T} {700, 1680}

\bibitem[\protect\citeauthoryear{{Tanvir}, {Kruehler}, {Wiersema}, {Xu},
  {Malesani}, {Milvang-Jensen}  \& {Fynbo}}{{Tanvir} et~al.}{2016}]{Tanvir16}
{Tanvir} N.~R.,  {Kruehler} T.,  {Wiersema} K.,  {Xu} D.,  {Malesani} D.,
  {Milvang-Jensen} B.,   {Fynbo} J.~P.~U.,  2016, GRB Coordinates Network,
  \href {http://adsabs.harvard.edu/abs/2016GCN..20321...1T} {20321}

\bibitem[\protect\citeauthoryear{{Toy} et~al.,}{{Toy} et~al.}{2016}]{Toy16}
{Toy} V.~L.,  et~al., 2016, \mn@doi [\apj] {10.3847/0004-637X/818/1/79}, \href
  {http://adsabs.harvard.edu/abs/2016ApJ...818...79T} {818, 79}

\bibitem[\protect\citeauthoryear{{Ugliano}, {Janka}, {Marek}  \&
  {Arcones}}{{Ugliano} et~al.}{2012}]{Ugliano12}
{Ugliano} M.,  {Janka} H.-T.,  {Marek} A.,   {Arcones} A.,  2012, \mn@doi
  [\apj] {10.1088/0004-637X/757/1/69}, \href
  {http://adsabs.harvard.edu/abs/2012ApJ...757...69U} {757, 69}

\bibitem[\protect\citeauthoryear{{Uzdensky} \& {MacFadyen}}{{Uzdensky} \&
  {MacFadyen}}{2007}]{Uzdensky07}
{Uzdensky} D.~A.,  {MacFadyen} A.~I.,  2007, \mn@doi [\apj] {10.1086/521322},
  \href {http://adsabs.harvard.edu/abs/2007ApJ...669..546U} {669, 546}

\bibitem[\protect\citeauthoryear{{Woosley}}{{Woosley}}{1993}]{Woosley93}
{Woosley} S.~E.,  1993, \mn@doi [\apj] {10.1086/172359}, \href
  {http://adsabs.harvard.edu/abs/1993ApJ...405..273W} {405, 273}

\bibitem[\protect\citeauthoryear{{Woosley}}{{Woosley}}{2010}]{Woosley10}
{Woosley} S.~E.,  2010, \mn@doi [\apjl] {10.1088/2041-8205/719/2/L204}, \href
  {http://adsabs.harvard.edu/abs/2010ApJ...719L.204W} {719, L204}

\bibitem[\protect\citeauthoryear{{Woosley}}{{Woosley}}{2017}]{Woosley17}
{Woosley} S.~E.,  2017, \mn@doi [\apj] {10.3847/1538-4357/836/2/244}, \href
  {http://adsabs.harvard.edu/abs/2017ApJ...836..244W} {836, 244}

\bibitem[\protect\citeauthoryear{{Woosley} \& {Bloom}}{{Woosley} \&
  {Bloom}}{2006}]{Woosley06}
{Woosley} S.~E.,  {Bloom} J.~S.,  2006, \mn@doi [\araa]
  {10.1146/annurev.astro.43.072103.150558}, \href
  {http://adsabs.harvard.edu/abs/2006ARA%26A..44..507W} {44, 507}

\bibitem[\protect\citeauthoryear{{Zhang}, {Woosley}  \& {Heger}}{{Zhang}
  et~al.}{2004}]{Zhang04}
{Zhang} W.,  {Woosley} S.~E.,   {Heger} A.,  2004, \mn@doi [\apj]
  {10.1086/386300}, \href {http://adsabs.harvard.edu/abs/2004ApJ...608..365Z}
  {608, 365}

\makeatother
\end{thebibliography}



\appendix
\begin{table*}[htb]
\begin{minipage}[t]{\textwidth}
\renewcommand{\footnoterule}{}
\centering
\caption{Secondary standard stars in the field of view of FORS images.}
\begin{tabular}{r|c|cccc}
\hline
\# & $\alpha$, $\delta$ (J2000) (deg) &  $B$\,(mag)  &  $V$\,(mag)  &  $R$\,(mag) &  $I$\,(mag)  \\
\hline
1  & 91.67230 $-$26.80637 & 20.671(11) & 19.329(06) &  18.518(06) & 17.678(06)  \\ 
2  & 91.67459 $-$26.8301  & 18.380(03) & 17.826(05) &  17.454(04) & 17.086(05)  \\ 
3  & 91.69784 $-$26.81922 & 21.895(27) & 21.320(27) &  20.925(25) & 20.557(41)  \\ 
4  & 91.71324 $-$26.83015 & 21.810(27) & 20.723(14)  &  20.042(12) & 19.407(17) \\ 
5  & 91.71604 $-$26.77713 & 22.030(45) & 21.416(45)  &20.937(49)& 20.425(63)    \\ 
6  & 91.72399 $-$26.82654 & 19.903(06) & 19.334(06) & 18.939(07) & 18.540(10)  \\ 
7  & 91.73697 $-$26.77008 & 21.223(17) & 19.918(09) &  19.152(07) & 18.397(10)  \\ 
8  & 91.73742 $-$26.78597 & 22.547(41) & 21.027(20)  &  20.010(11)& 18.579(12)  \\ 
9  & 91.74192 $-$26.79532 & 19.325(05) & 18.516(05)  &  18.046(04)& 17.583(09)  \\ 
10 & 91.74498 $-$26.78878 & 18.638(13) & 17.766(11)  &  17.242(07)& 16.750(07)  \\ 
11 & 91.75173 $-$26.80814 & 22.006(31) & 21.576(32)  &  21.204(30)& 20.866(57)  \\
12 & 91.75232 $-$26.82249 & 21.161(16) & 20.674(15)  &  20.332(16)& 19.944(28)  \\ 
\hline
\end{tabular}
\footnote{Numbers in parentheses give the photometric 1$\sigma$ statistical uncertainty of the secondary standards in units of 10\,millimag.}
\label{table:std}
\end{minipage}
\end{table*}

\section{Constraints on the re-energised blast wave}

The blast wave that re-energises the afterglow before it started to be observed is assumed to have a form $L(t) \propto t^{-q}$. We determined $q$ following the formalism of  \citep{Dai01} for afterglows produced  in uniform media by relativistic electron distributions with a flat spectrum ($p < 2$), and by comparing with the observed decay indices at optical and X-ray wavelengths. The condition that must be satisfied is $q < 1$.  

We have

$E_{\rm tot} \approx E_{\rm inj} \propto t^{1 - q},$

$\Gamma^2 R^3 \propto t^{1 - q},$

\noindent
but also $\Gamma^2 R^3 \propto R^{1 - q} \Gamma^{2(q - 1)}$, because $t \propto R/\Gamma^2$;

\noindent
then

$\Gamma \propto R^{-(2+q)/(4-2q)} \propto t^{-(2+q)/8}$, so that $R \propto t^{(2-q)/4}$.

\noindent
We have $B \propto \Gamma$ and

$\gamma_m \propto \Gamma^{1/(p-1)} B^{-(p-2)/[2(p-1)]} \propto \Gamma^{(4-p)/[2(p-1)]}$,

$\gamma_c \propto \Gamma^{-1} B^{-2} t^{-1}$.

\noindent
For the case of a blast wave propagating in a homogeneous medium,

$\nu_m \propto \Gamma \gamma_m^2 B \propto \Gamma^{(p+2)/(p-1)} \propto t^{-(2+q)(p+2)/[8(p-1)]},$

$\nu_c \propto \Gamma \gamma_c^2 B \propto \Gamma^{-4} t^{-2} = t^{(q-2)/2},$ and

$F_{\rm max} \propto N_e \Gamma B \propto N_e \Gamma^2 \propto \Gamma^2 R^3 \propto t^{1 - q}$ .

\noindent
We then have, for $\nu_m < \nu < \nu_c$,

$F_{\nu} \propto F_{\rm max} \nu_m^{(p-1)/2} \propto t^{1 - q} t^{-(2+q)(p+2)/16}$.

\noindent
This expression for $F_{\nu}$ must be equivalent to $t^{-\alpha_{\rm opt}}$, 
from which we derive $\alpha_{\rm opt} = (2+q)(p+2)/16 + q - 1$, which in turn gives $q = 0.85$ for $\alpha_{\rm opt} \approx 0.5$ (observed optical decay index) and $p = 1.66$.

For $\nu > \nu_c$, we have 

$F_{\nu} \propto F_{\rm max} \nu_m^{(p-1)/2} \nu_c^{1/2} \propto t^{-\alpha_{\rm opt}} t^{-(2-q)/4} \propto t^{-\alpha_{\rm X}}$,

\noindent
which, for $q = 0.85$, gives $\alpha_{\rm X} \approx 0.8$, as observed.


\bsp	
\label{lastpage}
\end{document}